\documentclass[%draft % uncomment if draft mode is needed
               ]{jetp} % jetp.cls

\usepackage{epsfig,color}
\usepackage{bm}% bold math
\usepackage{graphicx}
\usepackage{changes}
\twocolumn % закомментировать, если не нужен двухколоночный вариант
\begin{document}
%\English

\def\TitleSkip{\vskip 5mm}
\def\AuthorsSkip{\bigskip}
\def\AffilSkip{\medskip}
\def\AffiliationsSkip{\medskip}
\def\DatesSkip{\bigskip}
\def\DOISkip{\vskip -5mm}
\def\LFVS{\vskip -5mm}
\def\ELFVS{\vspace*{0mm}}

\title{Electronic structure of the $CuO$ monolayer in the paramagnetic phase considering the Coulomb interactions}

\setaffiliation1{Kirensky Institute of Physics, Federal Research Center KSC SB RAS \\ 660036, Krasnoyarsk, Russia} %
\setaffiliation2{Institute of Electrophysics of the Ural Branch of the Russian Academy of Sciences\\ 620016, Ekaterinburg, Russia}
\setaffiliation3{Р.N. Lebedev Physical Institute of the Russian Academy of Sciences\\ 119991, Moscow, Russia}
\setaffiliation4{Federal Research Center "Krasnoyarsk Science Center of the Siberian Branch of the Russian Academy of Sciences"\\ 660036, Krasnoyarsk, Russia}

\setauthor{I.~A.}{Makarov}{1}
\email{maki@iph.krasn.ru}

\setauthor{A.~A.}{Slobodchikov}{2}

\setauthor{I.~A.}{Nekrasov}{23}

\setauthor{Yu.~S.}{Orlov}{1}

\setauthor{L.~V.}{Begunovich}{4}

\setauthor{M.~M.}{Korshunov}{1}

\setauthor{S.~G.}{Ovchinnikov}{1}

\rtitle{Electronic structure of $CuO$ monolayer\dots}
\rauthor{I.~A.~Makarov, A.~A.~Slobodchikov, I.~A.~Nekrasov and others.}

\abstract{The electronic structure of the CuO monolayer is investigated taking into account the intra- and interatomic Coulomb interactions on copper and oxygen atoms. Local Coulomb interactions and covalence effects are treated exactly when constructing quasiparticle excitations using the generalized tight-binding method (GTB). The electronic system is described in the hole representation within the eight-band $p-d$ model including long-range hoppings up to four nearest neighbors with parameters obtained from calculations using the density functional theory. Using the multiband Hubbard model, we calculated the dependencies of the band structure of quasiparticle excitations, Fermi surfaces, constant energy maps at the top of the valence band, band gap values and contributions of different orbitals to states at the top of the valence band in the regime of strong intra-atomic Coulomb interactions on copper for different values of the intra-atomic interaction on oxygen $U_p$ and the interatomic copper-oxygen interaction $V_{pd}$. It is shown that the system goes from the insulating state with $d$-states at the top of the valence band to the metallic state in which the main contribution to low-energy excitations is made by oxygen orbitals depending on the values of the parameters $U_p$ and $V_{pd}$.}

\maketitle
%\onecolumn
%\begin{multicols}{2}
\section{Introduction}

Among copper compounds, the class of HTSC cuprates attracts the most attention due to the presence of such properties as high-temperature superconductivity, pseudogap state, spin and charge ordering. The results of a huge number of experimental and theoretical studies of HTSC cuprates including films with a thickness of one unit cell~\cite{Bozovic2001,Bozovic2003,Bollinger2011,Gozar2016,Bollinger2016,Kim2023} and CuO$_2$ monolayers~\cite{Logvenov2009,Zhong2016} allow us to confidently claim that their common structural element - the CuO$_2$ plane - plays the key role in the formation of unusual properties \cite{Dagotto1994,Lee2006}. The discovery of superconductivity in the nickelates Nd$_{1-x}$Sr$_x$NiO$_2$, Nd$_{1-x}$Eu$_x$NiO$_2$ and LaNiO$_3$/La$_{0.7}$Sr$_{0.3}$MnO$_3$ superlattice~\cite{Anisimov1999,Zhang2017,Zhou2018,Li2019,Wei2023}, the presence of superconducting gap features in Sr$_2$IrO$_4$~\cite{Yan2015,Kim2016} confirm that the unusual superconductivity is caused by a certain type of electronic and crystal structure characteristic of HTSC cuprates and these compounds. In the square lattice of CuO$_2$ layer, the nearest copper atoms with four filled $3d$-orbitals and one half-filled ${d_{{x^2} - {y^2}}}$-orbital are separated by oxygen atoms with completely filled $2p$-orbitals in undoped cuprates. The long-range magnetic order is destroyed when doped holes are placed on oxygen orbitals, and the emergence of a superconducting state becomes possible, as expected, due to the superexchange interaction and spin fluctuations. In recent works~\cite{Kano2017,Yin2017,Kvashnin2019}, a new form of copper-oxygen planes in the form of a CuO monolayer was synthesized on graphene layers, in its pores and in a free state. The CuO monolayer and the CuO$_2$ plane in HTSC cuprates have both similarities and significant differences. The CuO monolayer is a two-dimensional, one-atom-thick, copper-oxygen plane with a square lattice. There is one hole in the unit cell of undoped CuO, the copper atoms have the $3d^9$ configuration, and the oxygen atoms have $2p^6$ configuration, i.e. the same as in HTSC cuprates. Unlike the CuO$_2$ layer, the oxygen atoms in the CuO monolayer separate not the nearest, but the next-to-nearest neighbors of copper atoms in the $\left[ {110} \right]$ direction. This will significantly affect the dispersion, magnetic correlations and the degree of localization/delocalization of charge carriers on the valence shell of the copper atoms. Unlike HTSC cuprates, the CuO monolayers have been studied very poorly so far, so it is unclear what the electron and magnetic systems of these compounds are, whether two-dimensional superconductivity is possible in them, and whether they have other unique properties of HTSC cuprates. A study of the electronic structure of the CuO monolayer and its comparison with the characteristics of HTSC cuprates should shed light on the nature of low-energy excitations in these systems and their relationship with the superconducting state.

Initially, a cubic structure similar to the NaCl structure and a correlated antiferromagnetic insulator state with high Neel temperatures ($700-800$ K) similar to other 3d transition metal monoxides (MnO, FeO, CoO, NiO)~\cite{Mattheiss1972,Terakura1984,Harrison2007,Zaanen1987} were predicted for CuO. It turned out that the bulk CuO has a low-symmetry monoclinic structure~\cite{Asbrink1970}, and the Neel temperature is only about $200$~K. In the work~\cite{Siemons2009}, it was possible to obtain a tetragonal CuO structure (T-CuO) elongated along the $z$ axis using epitaxial growth on the SrTiO$_3$ substrate. In the works~\cite{Siemons2009,Samal2014}, it was shown that the tetragonal CuO structure becomes stable in the form of thin films with the thickness of several unit cells. T-CuO consists of CuO planes, neighboring planes along the $z$ axis are shifted by half of a unit cell in the $x$ or $y$ direction, and each copper atom with the nearest oxygen atoms forms CuO$_6$ octahedra~\cite{Samal2014}. Since the structure with an elongated CuO$_6$ octahedron has the absolute minimum energy~\cite{Peralta2009,Chen2009}, the only hole on copper atoms will be located in the ${d_{{x^2} - {y^2}}}$-orbital which is confirmed by XAS experiments~\cite{Samal2014}. \textit{Ab initio} calculations within the local density functional theory~\cite{Filippetti2003} show that tetragonal AF-II CuO with an elongated octahedron should be an insulator with the band gap of $1.1$ eV~\cite{Peralta2009} while calculations using the hybrid density functional theory based on the Heyd-Scuseria-Ernzerhof method~\cite{Chen2009} predict the band gap of $2.7$ eV. The insulating state was also obtained in the LDA+U approximation with the Coulomb repulsion value exceeding $6.52$ eV~\cite{Grant2008}. In the work~\cite{Moser2014}, the band structure of the valence band and the constant energy maps at the top of the valence band simulating the Fermi contour were investigated for the T-CuO layer with the thickness of 6 unit cells using ARPES (angle-resolved photoemission spectroscopy). The lower limit of the band gap was estimated to be $2.35$ eV. The band structure resembles the dispersion in cuprates rotated by $45$ degrees in the Brillouin zone, the state of the valence band with the maximum energy is in the direction $\left( {0,0} \right) - \left( {\pi ,0} \right)$ (here and below, the values of the wave vectors will be given in units of $\frac{1}{a}$, where $a$ is the lattice constant)~\cite{Moser2014}. It can be concluded from the topology of the constant energy maps that the CuO layers are formed as two sublattices of copper atoms, one of which is located above and the other one below the plane of oxygen atoms, and neighboring copper atoms belong to different sublattices.

The metallic state with the band of ${d_{{x^2} - {y^2}}}$-, ${p_x}$-, ${p_y}$-orbitals at the Fermi level and the band of ${d_{xz}}$-, ${d_{yz}}$-, and ${p_z}$-orbitals located very close to it at the point $\left( { \pi,0} \right)$~\cite{Slobodchikov2023} was obtained for the CuO monolayer using the density functional theory (DFT) calculations in the LDA approximation. Since the strong electron correlations (SEC) can play an important role in CuO monolayers, as well as in cuprates, they must be taken into account in the calculations. Two different versions of the band structure were obtained within the framework of the DFT+U approach~\cite{Kano2017,Yin2017}. In the work~\cite{Kano2017}, the difference $U-J$ was estimated to be $8.5$~eV based on the correct reproduction of the band gap in the bulk monoclinic CuO. The insulating state of the CuO monolayer was obtained within the LSDA+U approximation~\cite{Dudarev1998}, and it was also found that the indirect band gap has a value of $2.7$~eV, the valence band maximum is at the point $\left( { \pi,0} \right)$, the main contribution to the states at the bottom of the conductivity band is made by copper orbitals, and to the states at the top of the valence band - by the oxygen orbitals~\cite{Kano2017}. The band structure with the indirect band gap of size $3.37$~eV and the valence band top in the direction $\left( { 0,0} \right)$-$\left( { \pi,0} \right)$ was obtained in the DFT+U calculations with $U=6.52$~eV~\cite{Yin2017}.

The single-electron approach leads to incorrect results for systems with SEC. To correctly take into account SEC and multiparticle effects, the GTB (generalized tight-binding) method that is a cluster form of the perturbation theory in terms of Hubbard operators was developed~\cite{Ovchinnikov89,Gavrichkov00,Korshunov05,OvchinnikovValkov}. In this paper, the GTB method is generalized to describe the electronic structure of the CuO monolayer exactly considering the strong local interactions, such as the Coulomb interaction and copper-oxygen hybridization, while constructing local multiparticle states the quasiparticle excitations between which will form the electronic structure. The electronic system of the CuO monolayer is described within the eight-band $p-d$ model including the Coulomb intra-atomic intra-orbital ${U_d}$, inter-orbital ${V_d}$, Hund's ${J_d}$ interactions on copper atoms, intra-atomic interactions on oxygen atoms ${U_p}$ and interatomic interactions between copper and oxygen atoms ${V_{pd}}$. The values of the parameters ${U_p}$ and ${V_{pd}}$ will be discussed below in Section 2. The electronic structure of quasiparticle excitations is calculated within the multi-band Hubbard model using the equations of motion for Green's functions in the Hubbard I approximation.

The paper consists of five sections. Section 2 contains the crystal lattice, basis atomic orbitals, and the Hamiltonian of the eight-band $p-d$ model in the electron representation. The transition to the hole representation for the Hamiltonian and the generalization of the GTB method for describing the electronic structure of the CuO monolayer are considered in Section 3. Section 4 presents the electronic structure without Coulomb interactions. Section 5 describes the results on the electronic structure in the regime of strong Coulomb intra-atomic interaction on copper atoms and the influence of the Coulomb parameters ${U_p}$ and ${V_{pd}}$ on it. The conclusion contains the main results of the paper.

\begin{figure}
\fig[1]{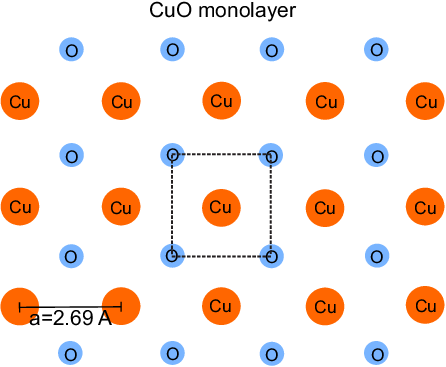}{}{}{Crystal structure of the CuO monolayer. The dashed lines indicate the boundaries of the unit cell.}
\end{figure}

\section{Hamiltonian of the eight-band $p-d$ model in the electron representation}
%\end{multicols}{2}
%\begin{multicols}{1}
%\onecolumn%\begin{multicols}{2}
%\twocolumn

%\vspace{-7mm}
%\hspace{2mm}
%\end{multicols}{1}
%\onecolumn\begin{multicols}{2}
A CuO monolayer is a two-dimensional plane with a square lattice consisting of copper atoms (orbital number ${l_{Cu}} = 2$) and oxygen atoms (orbital number ${l_{O}} = 1$), the lattice parameters are $a=b=2.69$~$\textrm{\AA}$~\cite{Yin2017}, see Fig.~1. The distance between the nearest copper and oxygen atoms is ${R_{Cu - O}} = 1.9$~$\textrm{\AA}$, that is, very close to the analogous value in the cuprates La$_{2-x}$Sr$_x$CuO$_4$ and Bi$_2$Sr$_2$CaCu$_2$O$_8$. In the CuO monolayer, each copper atom is surrounded by four oxygen atoms (${z_{dp}} = 4$), and each oxygen atom is surrounded by four copper atoms (${z_{pd}} = 4$). The unit cell is a square with one copper atom in the center and four oxygen atoms in the corners (Fig.~1). In the CuO structure, copper ions have the charge $2+$ and the electron configuration of valence orbitals $3d^9$, oxygen ions have the charge of $2-$ and the electron configuration $2p^6$. We choose eight electron atomic orbitals as a basis of the multiband $p-d$ model for the CuO monolayer: five $3d$ orbitals on the copper atom ${d_{{x^2} - {y^2}}}$, ${d_{3{z^2} - {r^2}}}$, ${d_{xy}}$, ${d_{xz}}$, ${d_{yz}}$, and three $2p$ oxygen orbitals ${p_x}$, ${p_y}$, ${p_z}$. They are shown in Fig.~2a,b, where the signs of the phases of the wave functions obtained within the density functional theory are given. The highest-energy electron $d$-orbital of copper is the ${d_{{x^2} - {y^2}}}$-orbital (Table~1, Fig.~3), and it is the ${d_{{x^2} - {y^2}}}$-orbital that will be half filled with electrons in the stoichiometric composition of the CuO monolayer.

\onecolumn\begin{multicols}{2}

In the electron representation, there are 15 electrons in each unit cell, the occupation of the atomic orbitals by them is shown in Fig.~3. The energy values are given below in Fig.~3 and in Table~1, as well as the modules of the hopping integrals in Tables~2-4 in Appendix 1, were obtained from the DFT calculations~\cite{Slobodchikov2023}. In the hole representation, the electron configuration $3d^9$ corresponds to the state with one hole against the background of completely filled electron shells. The hole can occupy other copper $d$-orbitals and oxygen $p$-orbitals due to the overlap of copper and oxygen orbitals.

The Hamiltonian of the eight-band $p-d$ model has the form:
\begin{eqnarray}
H &=& \sum\limits_{\mathbf{r} \lambda \sigma } {\left( {{\varepsilon _{\mathbf{r}\lambda} } - \mu } \right){n_{\mathbf{r}\lambda \sigma }}}  +\nonumber\\
&+& \sum\limits_{\mathbf{r} \mathbf{r}' \lambda \lambda'} {{t_{\lambda \lambda '}}\left( {{{\bf{R}}_{\lambda \lambda '}}} \right)c_{\mathbf{r}\lambda \sigma }^\dag {c_{\mathbf{r}' \lambda '\sigma }}} + {H_U}, \nonumber\\
{H_U} &=& \sum\limits_{\mathbf{r} \lambda}  {{U_\lambda }{n_{\mathbf{r}\lambda  \downarrow }}{n_{\mathbf{r}\lambda  \uparrow }}}  + \frac{1}{2}\sum\limits_{\mathbf{r} \lambda  \ne \lambda' \sigma \sigma'} {{V_{\lambda \lambda '}}{n_{\mathbf{r}\lambda \sigma }}{n_{\mathbf{r}\lambda '\sigma '}}} - \nonumber\\
&-& \frac{1}{2}\sum\limits_{\mathbf{r} \lambda \ne \lambda' \sigma \sigma'} {{J_{\lambda \lambda '}}c_{\mathbf{r}\lambda \sigma }^\dag {c_{\mathbf{r}\lambda \sigma '}}c_{\mathbf{r}\lambda '\sigma '}^\dag {c_{\mathbf{r}\lambda '\sigma }}}.
\label{eq:Ham_el}
\end{eqnarray}
%\endlongformula
Here the indices $\lambda$, $\lambda '$ run over the values of all atomic $d$-orbitals of copper and $p$-orbitals of oxygen in the CuO monolayer, ${c_{\mathbf{r}\lambda \sigma }}$ is the electron annihilation operator with the spin projection $\sigma$ on the site $\mathbf{r}$ with the orbital index $\lambda$, ${c_{\mathbf{r}\lambda \sigma }}={d_{\mathbf{r}\lambda \sigma}}$ or ${p_{\mathbf{r}\lambda \sigma}}$, ${{n_{\mathbf{r}\lambda \sigma }}}={c_{\mathbf{r}\lambda \sigma }^\dag {c_{\mathbf{r}\lambda \sigma }}}$. The parameter ${\varepsilon _{\mathbf{r}\lambda }}$ is the on-site electron energy for the ${\lambda }$ orbital, ${\varepsilon _{\mathbf{r}\lambda} } = {\varepsilon _{d \lambda}}$ for copper $d$-orbitals and ${\varepsilon _{\mathbf{r}\lambda} } = {\varepsilon _{p \lambda}}$ for oxygen $p$-orbitals. ${t_{\lambda \lambda '}}\left( {{{\bf{R}}_{\lambda \lambda '}}} \right)$ is the hopping integral between the orbitals $\lambda$ and $\lambda '$, the vector ${\bf{R}}_{\lambda \lambda '} = \mathbf{r} - \mathbf{r}'$ connects atoms at the sites $\mathbf{r}$ and $\mathbf{r}'$, where these orbitals are located. The Cu-Cu, O-O hopping integrals between five nearest neighbors and Cu-O hopping integrals between four nearest neighbors are considered (Tables 2-4 in Appendix 1). The intra-atomic Coulomb interaction parameter ${U_\lambda }$ is equal to ${U_d }$ for copper orbitals and $U_p$ for oxygen orbitals. The parameter ${V_{\lambda \lambda '}}$ is the intra-atomic interorbital Coulomb interaction if the orbitals $\lambda$ and $\lambda '$ are on the same atom (${V_{\lambda \lambda '}}={V_d }$ for Cu and $V_p=U_p$~--- for O), and the interatomic interaction $V_{pd}$ if one of the orbitals $\lambda$, $\lambda '$ is on Cu, and the other~-- on O, where the copper-oxygen interaction $V_{pd}$ is considered only between the nearest neighbors. The exchange interaction ${J_{\lambda \lambda '}}$ is taken into account only between the $d$-orbitals inside a copper atom (${J_{\lambda \lambda '}}=J_d$, $J_p=0$). The difference in energies between the interorbital Coulomb interactions of two electrons located on different pairs of orbitals of one atom is small compared to their absolute value. For example, this difference for the Coulomb parameters ${V_{{d_{{x^2} - {y^2}}}{d_{3{z^2} - {r^2}}}}}$ and ${V_{{d_{{x^2} - {y^2}}}{d_{xy}}}}$ is $4.8\%$ of the value of ${V_{{d_{{x^2} - {y^2}}}{d_{xy}}}}$. Therefore, we will use the approximation in which it is assumed that the interorbital interaction between two electrons located in different pairs of orbitals of the same shells of the same or different atoms have the same value when considering the Coulomb $V_d$, $U_p$, $V_{pd}$ and exchange $J_d$ interactions. It is also assumed that the values of intra-atomic Coulomb interactions on copper atoms ${U_d}$, ${V_d}$, ${J_d}$ are close to their values in cuprates~\cite{Schluter88,Schluter89,Hybertsen89,Hybertsen90,Mahan90,Grant1992,Gunnarsson1989,Anisimov1991,Anisimov1992} and to those calculated in the LDA+U approach for the CuO monolayer~\cite{Kano2017}, therefore we further adopt the following values (except for specially highlighted cases): ${U_d=9}$, ${V_d=7}$, ${J_d=1}$~eV. Also, the values of the parameters ${U_p}$, ${V_{pd}}$ will be taken close to the values in cuprates (${U_p}=3$, ${V_{pd}}=1.5$~eV). Then their values will be varied in a small range to study possible changes in the electronic structure of the CuO monolayer.

\longformula

\begin{figure}
\fig[2]{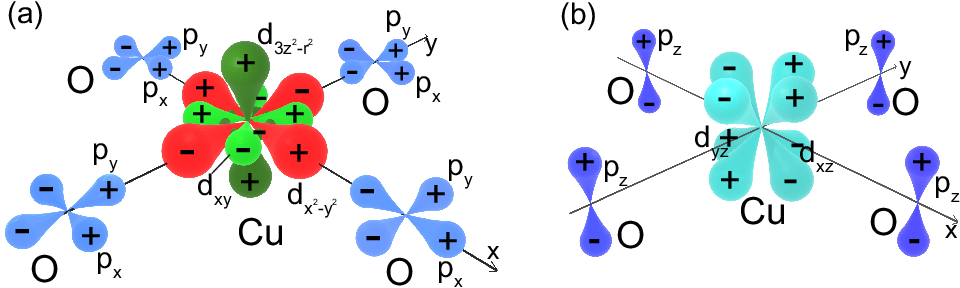}{}{}{(a) Basis atomic orbitals ${d_{{x^2} - {y^2}}}$, ${d_{xy}}$, ${d_{3{z^2} - {r^2}}}$ of copper and ${p_x}$, ${p_y}$ of oxygen for the multiband $p-d$ model in the unit cell of CuO. (b) Basis atomic orbitals ${d_{xz}}$, ${d_{yz}}$ of copper and ${p_z}$ of oxygen in the CuO unit cell. The signs of the wave function phases are shown for each orbital.}
\end{figure}

\begin{figure}
\fig[3]{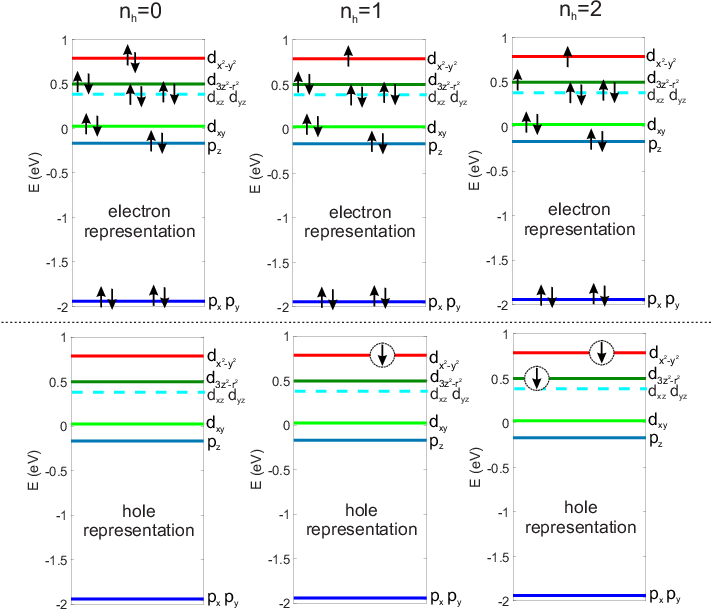}{90cm}{90cm}{Occupation of the copper and oxygen atomic orbitals of the CuO cluster in the electron and hole representations. The single-hole state corresponds to the stoichiometric composition of the CuO monolayer.}
\end{figure}

\begin{table}
\tabl[1]{{\sffamily\small\bfseries } The values of on-site energies in the Hamiltonian of the eight-band $p-d$ model obtained from the DFT calculations~\cite{Slobodchikov2023} (in eV)}
\begin{center}
\begin{tabular}{|c|c|c|c|c|c|}
\hline
\rule{0mm}{5mm} ${\varepsilon _{{d_{{x^2} - {y^2}}}}}$ & ${\varepsilon _{d_{3{z^2} - {r^2}}}}$ & ${\varepsilon _{d_{xy}}}$ & ${\varepsilon _{d_{xz}}}$ (${\varepsilon _{d_{yz}}}$) & ${\varepsilon _{p_x}}$ (${\varepsilon _{p_y}}$) & ${\varepsilon _{p_z}}$ \\
\hline
\rule{0mm}{5mm}
$0.787$ & $0.498$ & $0.024$ & $0.3830$ & $-1.942$ & $-0.168$ \\
\hline
\end{tabular}
\end{center}
\end{table}

\endlongformula

\section{GTB method for $CuO$ monolayer}
In the GTB method, local multiparticle states and their energies are calculated using the exact diagonalization of the Hamiltonian of one of the identical clusters, into the set of which the entire crystal lattice of the compound under consideration is divided. Fermi-type excitations between these states are described by Hubbard operators. Then, a complete or effective multiband Hubbard model for quasiparticle excitations is constructed.

The minimum cluster size for a CuO monolayer is one unit cell which includes $15$ electrons on ten orbitals of the highest-energy shells in the undoped compound. The basis for constructing elementary Fermi type quasiparticle excitations consists of multiparticle states with 14, 15, and 16 electrons (Fig. 3). The consideration of such a number of basis states and quasiparticle excitations requires a lot of time and computer resources. On the other hand, the indicated multiparticle states are two-hole, single-hole and zero-hole states (Fig.~3) in the hole representation (against the background of completely filled electron shells $d^{10}p^6$), and the number of excitations between them is significantly reduced. Therefore, we should switch to the hole representation for convenience of description.

\subsection{Hole representation of the $p-d$ model Hamiltonian}
The ground state of electrons in the second quantization representation is the vacuum state ${\left| {{{...,0}_{\lambda ' \downarrow }}{{,0}_{\lambda ' \uparrow }}{{,...,0}_{\lambda \downarrow }}{{,0}_{\lambda \uparrow }},...} \right\rangle _e}$ with zero electrons on each orbital $\lambda $ in the CuO monolayer. To transit to the hole representation, we take the vacuum state of holes to be the state with completely filled electron states ${\left| {FB} \right\rangle _e}$ ("full band")~\cite{Mahan2000}:
\begin{eqnarray}
{\left| {FB} \right\rangle _e} &=& {\left| {{{...,1}_{\lambda ' \downarrow }}{{,1}_{\lambda ' \uparrow }}{{,...,1}_{\lambda  \downarrow }}{{,1}_{\lambda  \uparrow }},...} \right\rangle _e} = \nonumber\\
&=& {\left| {{{...,0}_{\lambda ' \downarrow }}{{,0}_{\lambda ' \uparrow }}{{,...,0}_{\lambda  \downarrow }}{{,0}_{\lambda  \uparrow }},...} \right\rangle _h} = {\left| 0 \right\rangle _h}.
\end{eqnarray}
The relations between the creation and annihilation operators in the electron and hole representations are defined as follows~\cite{Mahan2000}:
\begin{equation}
h_{\lambda \bar \sigma }^\dag {\left| 0 \right\rangle _h} = {c_{\lambda \sigma }}{\left| {FB} \right\rangle _e}, \;\;\;
{h_{\lambda \bar \sigma }}{\left| 0 \right\rangle _h} = c_{\lambda \sigma }^\dag {\left| {FB} \right\rangle _e}
\end{equation}
where ${{h_{\lambda \sigma }}}$ ($h_{\lambda \bar \sigma }^\dag$) is the hole annihilation (creation) operator. Then the Hamiltonian~(\ref{eq:Ham_el}) in the hole representation takes the form
%\longformula
\begin{eqnarray}
{H^{\left( h \right)}} &=& {\varepsilon _0} - \sum\limits_{\mathbf{r} \lambda \sigma } {\left( {{\varepsilon _{\mathbf{r}\lambda} }
+ {{\tilde U}_{\mathbf{r}\lambda} }\, - \mu } \right)h_{\mathbf{r}\lambda \sigma }^\dag {h_{\mathbf{r}\lambda \sigma }}}  - \nonumber\\
&-&\sum\limits_{\mathbf{r} \mathbf{r}' \lambda \lambda' \sigma} {{t_{\lambda \lambda '}}\left( {{{\bf{R}}_{\lambda \lambda '}}} \right)h_{\mathbf{r}\lambda \sigma }^\dag {h_{\mathbf{r}' \lambda '\sigma }}}  + H_U^{\left( h \right)}
\label{eq:Ham_hole}
\end{eqnarray}
%\endlongformula
Here ${\varepsilon _0}$ is the energy of the copper $3d$-shells and oxygen $2p$-shells (vacuum state of holes) completely filled with electrons on all atoms of the CuO monolayer (given in Appendix~3). The coefficients before operators of the type ${h_{\mathbf{r}\lambda\sigma }^\dag {h_{\mathbf{r}\lambda\sigma }}}$ in the Hamiltonian (\ref{eq:Ham_hole}) corresponding to on-site energies differ from the similar coefficients in the Hamiltonian (\ref{eq:Ham_el}) in that they include Coulomb interactions ${{{\tilde U}_{\mathbf{r}\lambda} }}$ (${{{\tilde U}_{\mathbf{r}\lambda} }}={{{\tilde U}_d }}$ for each $d$-orbital, ${{{\tilde U}_{\mathbf{r}\lambda} }}={{{\tilde U}_p }}$ for each $p$-orbital), and there is a minus sign before the entire expression. The energy of a hole added to a vacuum hole state is the energy that is extracted from the system when an electron is removed from the completely filled electron configuration. That is, when we remove an electron, we also remove all interactions involving it. Therefore, the energy of the system decreases by $\varepsilon _{\mathbf{r}\lambda} ^{\left( h \right)} = {\varepsilon _{d\lambda }^{\left( h \right)}}= {\varepsilon _{d \lambda}} + {{{\tilde U}_d }}$, where ${{{\tilde U}_d }} = {U_d} + 4{l_{Cu}}\left( {{V_d} + {J_d}} \right) + 2\left( {2{l_O} + 1} \right){z_{dp}}{V_{pd}}$ is the sum of the energies of the intraorbital, interorbital, and interatomic interactions in which the given electron was involved, when a hole is added to a $d_{\lambda}$ orbital. When a hole is added to the $p_{\lambda}$-orbital, the correction to the on-site energy ${\varepsilon _{p \lambda}}$ from Coulomb interactions has the form ${{\tilde U}_p} = \left( {4{l_{Cu}} + 1} \right){U_p} + 2\left( {2{l_{Cu}} + 1} \right){z_{pd}}{V_{pd}}$, $\varepsilon _{\mathbf{r}\lambda} ^{\left( h \right)} ={\varepsilon _{p\lambda }^{\left( h \right)}}= {\varepsilon _{p \lambda}} + {{{\tilde U}_p }}$. The form of the Hamiltonian of Coulomb interactions $H_U^{\left( h \right)}$ in the hole representation completely coincides with the form of the Hamiltonian in the electron representation ${H_U}$ except that it uses hole creation and annihilation operators. In what follows, the creation and annihilation operators at the site ${\bf{f}}$
\begin{eqnarray}
&&{d_{{{x^2} - {y^2}}{\bf{f}}\sigma }}, {d_{{3{z^2} - {r^2}}{\bf{f}}\sigma }}, {d_{xy{\bf{f}}\sigma }}, \\\nonumber
&&{d_{xz{\bf{f}}\sigma }}, {d_{yz{\bf{f}}\sigma }}, {p_{x{\bf{f}}\sigma }}, {p_{y{\bf{f}}\sigma }}, {p_{z{\bf{f}}\sigma }}\nonumber
\end{eqnarray}
are for holes, not electrons.
\subsection{Hamiltonian of the $p-d$ model in the hole representation in terms of oxygen molecular orbitals}
We choose the CuO unit cell as an elementary cluster (Fig.~1). Since each oxygen atom belongs to four unit cells simultaneously, the electron states in the oxygen orbitals are not orthogonal in neighboring clusters. To orthogonalize the states in the oxygen orbitals ${p_x}$, ${p_y}$, ${p_z}$, we generalize the Shastry's procedure proposed for the two orbitals ${p_x}$ and ${p_y}$~\cite{Shastry89,Raimondi1996} to the case of three orbitals. The orthogonal transformation in $k$-space from the orbitals ${p_{x{\bf{k}}}}$, ${p_{y{\bf{k}}}}$, ${p_{z{\bf{k}}}}$ to the new basis of the molecular orbitals ${\alpha _{\bf{k}}}$, ${\beta _{\bf{k}}}$, ${\gamma _{\bf{k}}}$ is given in Appendix~1. The Hamiltonian~(\ref{eq:Ham_hole}) in terms of new oxygen molecular orbitals takes the following form:
\longformula
\begin{multline}
{H^{\left( h \right)}} = {\varepsilon _0} - \sum\limits_{{\bf{f}}\zeta \sigma } {\left[ {\left( {\varepsilon _{d\zeta }^{\left( h \right)} - \mu } \right)d_{\zeta {\bf{f}}\sigma }^\dag {d_{\zeta {\bf{f}}\sigma }} + \sum\limits_{{\bf{R}}\zeta '} {{t_{\zeta \zeta '}}\left( {\bf{R}} \right)d_{\zeta {\bf{f}}\sigma }^\dag {d_{\zeta '\left( {{\bf{f}} + {\bf{R}}} \right)\sigma }} + \sum\limits_{{\bf{R}}j} {\left( {\kappa _{\bf{R}}^{\left( {\zeta {\rho _j}} \right)}d_{\zeta {\bf{f}}\sigma }^\dag {\rho _{j\left( {{\bf{f}} + {\bf{R}}} \right)\sigma }} + h.c.} \right)} } } \right]}  - \\
 - \sum\limits_{{\bf{fR}}ij\sigma } {\left[ {{\delta _{ij}}\left( {\nu _{\bf{R}}^{\left( i \right)} - \mu } \right) + \left( {1 - {\delta _{ij}}} \right)\nu _{\bf{R}}^{\left( {ij} \right)}} \right]\rho _{i{\bf{f}}\sigma }^\dag {\rho _{j\left( {{\bf{f}} + {\bf{R}}} \right)\sigma }}}  + \sum\limits_{{\bf{ff'gg'}}} {\sum\limits_{ii'jj'} {{U_p}\Psi _{{\bf{fghr}}}^{{\rho _i}{\rho _{i'}}{\rho _j}{\rho _{j'}}}\rho _{i{\bf{f}}\sigma }^\dag {\rho _{i'{\bf{f '}}\sigma }}\rho _{j{\bf{g}}\sigma '}^\dag {\rho _{j'{\bf{g '}}\sigma '}}} }  + \\
 + \sum\limits_{{\bf{f}}\zeta \sigma } {\left\{ {\frac{1}{2}{U_d}n_{\zeta {\bf{f}}\sigma }^{\left( d \right)}n_{\zeta {\bf{f}}\sigma }^{\left( d \right)} + \frac{1}{2}\sum\limits_{\zeta '\sigma '} {\left[ {{V_d}n_{\zeta {\bf{f}}\sigma }^{\left( d \right)}n_{\zeta '{\bf{f}}\sigma '}^{\left( d \right)} - {J_d}d_{\zeta {\bf{f}}\sigma }^\dag {d_{\zeta {\bf{f}}\sigma '}}d_{\zeta '{\bf{f}}\sigma '}^\dag {d_{\zeta '{\bf{f}}\sigma }}} \right]}  + \sum\limits_{{\bf{gg'}}ij\sigma '} {{V_{pd}}\Phi _{{\bf{fgg'}}}^{{\rho _i}{\rho _j}}d_{\zeta {\bf{f}}\sigma }^\dag {d_{\zeta {\bf{f}}\sigma }}\rho _{i{\bf{g}}\sigma '}^\dag {\rho _{j{\bf{g'}}\sigma '}}} } \right\}} ,
\label{eq:Ham_mo}
\end{multline}
\begin{figure}
\fig[4]{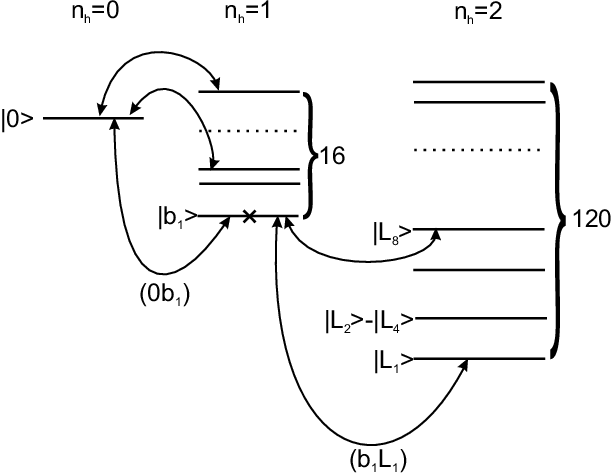}{}{}{Schematic representation of the eigenenergies of the CuO cluster with the number of holes ${n_h} = 0,1,2$ (horizontal lines) and the Fermi type quasiparticle excitations between them (curves with arrows). The cross indicates the only single-hole ground state filled in the stoichiometric composition of the CuO monolayer.}
\end{figure}
\endlongformula
where ${\bf{f}}$, ${\bf{g}}$ are the coordinates of the CuO clusters, ${\bf{R}}$ is the lattice translation vector. The index $\zeta$ corresponds to one of the five $d$-orbitals of copper; ${\rho _j}$ is the designation of one of the three molecular oxygen orbitals, ${\rho _1}\equiv\alpha$, ${\rho _2}\equiv\beta$, ${\rho _3}\equiv\gamma$. The definitions of the energy parameters ${\nu _{{\bf{fg}}}^{\left( i \right)}}$, ${\nu _{{\bf{fg}}}^{\left( {ij} \right)}}$, ${\kappa _{{\bf{fg}}}^{\left( {\lambda {\rho _j}} \right)}}$, ${\Phi _{{\bf{fgg'}}}^{{\rho _i}{\rho _j}}}$, $\Psi _{{\bf{ff'gg'}}}^{{\rho _i}{\rho _{i'}}{\rho _j}{\rho _{j'}}} $ are given in Appendix~4. We will further consider only the intracluster coefficients ${\Phi _{000}^{{\rho _i}{\rho _j}}}$ and $\Psi _{0000}^{{\rho _i}{\rho _{i'}}{\rho _j}{\rho _{j'}}} $ due to the smallness of the intercluster coefficients.

\subsection{Multiband Hubbard model and equations of motion for Green's functions}
The Hamiltonian~(\ref{eq:Ham_mo}) after the orthogonalization procedure can be represented as the sum of the Hamiltonian of intracluster interactions and the Hamiltonian of interactions of particles located in different clusters:
\begin{equation}
{H^{\left( h \right)}} = \sum\limits_{\bf{f}} {H_{c{\bf{f}}}^{\left( h \right)}}  + \sum\limits_{{\bf{fg}}} {H_{cc{\bf{fg}}}^{\left( h \right)}}.
\label{eq:divided_Ham}
\end{equation}
Next, we obtain the eigenstates $\left| {m} \right\rangle$ of an individual CuO cluster by solving the stationary Schrödinger equation:
\begin{equation}
H_{c{\bf{f}}}^{\left( h \right)}\left| m \right\rangle  = {\varepsilon _{m}}\left| m \right\rangle,
\label{eq:Schrod}
\end{equation}
where ${\varepsilon _{m}}$ is the energy of the many-particle eigenstate $\left| m \right\rangle$ of the CuO cluster. The exact diagonalization procedure is performed for a cluster with ${n_h} = 0,1,2$ holes in order to obtain the states into which the cluster transits from the stoichiometric composition with one hole under elementary Fermi excitations that create or annihilate one hole. The many-particle eigenstates of the cluster are one vacuum state with zero holes $\left| 0 \right\rangle $, sixteen single-hole states $\left| {b_{i\sigma }} \right\rangle $ (eight doublets), and $120$ two-hole states $\left| {{L_j}} \right\rangle $ (Fig.~4). In the undoped system, only the ground single-hole state is always filled, its components with the spin projections ${1 \mathord{\left/
{\vphantom {1 2}} \right.
\kern-\nulldelimiterspace} 2}$ and ${{ - 1} \mathord{\left/
{\vphantom {{ - 1} 2}} \right.
\kern-\nulldelimiterspace} 2}$ are degenerate in energy. The energy, orbital composition, symmetry, and probability density distribution between different orbitals depend on the Coulomb parameters. The Coulomb interactions $U_d$, $V_d$, $J_d$ increase the probability density to find a hole on copper $d$-orbitals in local states of the cluster, and the interactions $U_p$, $V_{pd}$ lead to an increase in the number of holes on oxygen $p$-orbitals. When the ratio between these types of Coulomb interactions changes, there will be a competition of states with different hole probability density distributions, as well as charge transfer between the $d$ and $p$-states. The reconstruction of local multiparticle eigenstates of the CuO cluster will affect the energy and orbital character of the quasiparticle excitations that form the electronic structure of the CuO monolayer. The quasiparticle excitation $\left( {m'm} \right)$ (Fig.~4) between the initial $\left| {m} \right\rangle $ and the final $\left| m' \right\rangle $ eigenstates of the CuO cluster at the site ${\bf{f}}$ with the number of fermions differing by one are described by the Hubbard operator $X_{\bf{f}}^{m'm} = \left| {m'} \right\rangle \left\langle {m} \right|$. Fermi operators for the annihilation (and creation) of holes in atomic or molecular orbitals are expressed in terms of Hubbard operators using the following relation:
\begin{equation}
h_{\lambda {\bf{f}}\sigma } = \sum\limits_{mm'} \gamma_{\lambda \sigma}^{m'm} X_{\bf{f}}^{m'm},
\label{eq:d_by_X}
\end{equation}
where $\gamma _{\lambda \sigma}^{m'm} = {\left\langle {m'} \right|{h_{\lambda {\bf{f}}\sigma }}\left| m \right\rangle}$.

To reproduce the complete electronic structure of the eight-band $p-d$ model in terms of quasiparticle excitations, it is necessary to consider the dispersion of all possible Fermi-type quasiparticles with $\left( {m'm} \right)$ transitions between all multiparticle eigenstates of the CuO cluster (Fig.~4). The basis of the multiband Hubbard model is formed from $1936$ quasiparticle excitations constructed on the basis of cluster multiparticle eigenstates. Some of these excitations have zero spectral weight, several due to zero matrix elements $\gamma_{\lambda \sigma}^{m'm}$, and others due to zero filling of the initial and final states. The Hamiltonian in terms of Hubbard $X$ operators has a standard form differing from the usual Hubbard model in the number of eigenstates that determine the size of the matrix of $X$ operators:
%\longformula
\begin{eqnarray}
H &=& \sum\limits_{{\mathbf{f}}m} {\left( {{\varepsilon _m} - \hat N\mu } \right)X_{\mathbf{f}}^{mm}}  + \nonumber\\
&+&\sum\limits_{{\mathbf{f}\mathbf{f}'}mm'n'n} {t^{mm'}_{n'n}(\mathbf{f}-\mathbf{f}')X_{\mathbf{f}}^{mm'}X_{\mathbf{f}'}^{n'n}},
\label{eq:H_Xabb}
\end{eqnarray}
%\endlongformula
where the indices $m$, $m'$, $n$, $n'$ run over the values of all zero-, single- and two-hole eigenstates of the CuO cluster. The hopping integrals $t^{mm'}_{n'n}(\mathbf{R}=\mathbf{f}-\mathbf{f}')$ between the quasiparticle excitations $\left( {mm'} \right)$ and $\left( {n'n} \right)$ are defined as follows:
\begin{eqnarray}
t^{mm'}_{n'n}(\mathbf{R}) &=& \sum\limits_{\zeta \zeta '\sigma } {{t_{\zeta \zeta'}}\left( {\bf{R}} \right)\gamma_{d\zeta \sigma }^{{mm'}*}\gamma_{{d\zeta '}\sigma }^{n'n}} + \nonumber\\
&+& \sum\limits_{ij\sigma } \nu _{\bf{R}}^{\left( {ij} \right)}\gamma_{{\rho _i}\sigma }^{{mm'}*}\gamma_{{\rho _j}\sigma}^{n'n} + \nonumber\\
&+& \sum\limits_{\zeta i \sigma } \left( {\kappa _{\bf{R}}^{\left( {\zeta {\rho_i}} \right)}\gamma_{d\zeta \sigma}^{{mm'}*}\gamma_{{\rho _i}\sigma }^{n'n}} + {{\kappa _{\bf{R}}^{\left( {\zeta {\rho_i}} \right)*}}}\gamma _{{\rho _i}\sigma }^{{mm'}*}\gamma _{d\zeta \sigma }^{n'n} \right).
\label{eq:hoppings_Xabb}
\end{eqnarray}

To obtain the electron spectrum, we will use the equation of motion method for the retarded electron Green's function ${G_{\lambda \lambda '\sigma }}\left( {{\bf{f}},{\bf{f'}};t} \right) = \left\langle {\left\langle {{{c_{\lambda {\bf{f}}\sigma }}\left( t \right)}}
 \mathrel{\left | {\vphantom {{{c_{\lambda {\bf{f}}\sigma }}\left( t \right)} {c_{\lambda '{\bf{f'}}\sigma }^\dag \left( 0 \right)}}}
 \right. \kern-\nulldelimiterspace}
 {{c_{\lambda '{\bf{f'}}\sigma }^\dag \left( 0 \right)}} \right\rangle } \right\rangle $. It is necessary to move from the electron Green's function to the hole Green's function of quasiparticle excitations:
\begin{multline}
%\left\langle {\left\langle {{{c_{\lambda {\bf{f}}\sigma }}\left( t \right)}}  \mathrel{\left | {\vphantom {{{c_{\lambda {\bf{f}}\sigma }}\left( t \right)} {c_{\lambda '{\bf{f'}}\sigma }^\dag \left( 0 \right)}}} \right. \kern-\nulldelimiterspace} {{c_{\lambda '{\bf{f'}}\sigma }^\dag \left( 0 \right)}} \right\rangle } \right\rangle
 {G_{\lambda \lambda '\sigma }}\left( {{\bf{f}},{\bf{f'}};t} \right) = \left\langle {\left\langle {{h_{\lambda {\bf{f}}\bar \sigma }^\dag \left( t \right)}}
 \mathrel{\left | {\vphantom {{h_{\lambda {\bf{f}}\bar \sigma }^\dag \left( t \right)} {{h_{\lambda '{\bf{f'}}\bar \sigma }}\left( 0 \right)}}}
 \right. \kern-\nulldelimiterspace}
 {{{h_{\lambda '{\bf{f'}}\bar \sigma }}\left( 0 \right)}} \right\rangle } \right\rangle  = \\
 = \sum\limits_{\lambda \lambda '} {\gamma _{\lambda \bar \sigma }^{mm' *} \gamma _{\lambda '\bar \sigma }^{n'n} \left\langle {\left\langle {{X_{\bf{f}}^{mm'}\left( t \right)}}
 \mathrel{\left | {\vphantom {{X_{\bf{f}}^{mm'}\left( t \right)} {X_{{\bf{f'}}}^{n'n}\left( 0 \right)}}}
 \right. \kern-\nulldelimiterspace}
 {{X_{{\bf{f'}}}^{n'n}\left( 0 \right)}} \right\rangle } \right\rangle } ,
\label{eq:Hole_GF_Xabb}
\end{multline}
where the number of holes in the state $m\left( n \right)$ is one more than in $m'\left( {n'} \right)$. Further, the equations of motion will be written for the Green's function ${D^{mm'}_{n'n}}\left( {{\bf{f}},{\bf{f'}};t} \right) = \left\langle {\left\langle {{X_{\bf{f}}^{mm'}\left( t \right)}}
\mathrel{\left | {\vphantom {{X_{\bf{f}}^{mm'}\left( t \right)} {X_{{\bf{f'}}}^{n'n}\left( 0 \right)}}}
\right. \kern-\nulldelimiterspace}
{{X_{{\bf{f'}}}^{n'n}\left( 0 \right)}} \right\rangle } \right\rangle $. The system of equations for the components of the matrix form of the Green's function $\hat D\left( {{\bf{f}},{\bf{f'}};\omega} \right)$ decoupled in the Hubbard I approximation has the form:
\begin{multline}
\omega {D^{mm'}_{n'n}}\left( {{\bf{f}},{\bf{f'}};{\omega}} \right) = {\delta _{mn}}{\delta _{m'n'}}F^{mm'} - \\
 - \Omega \left( {m'm} \right){D^{mm'}_{n'n}}\left( {{\bf{f}},{\bf{f'}};{\omega}} \right) - \\
 - \sum\limits_{pq} F^{pq} T^{mm'}_{pq}\left( {{\bf{f}},{\bf{f'}}} \right){D^{pq}_{n'n}}\left( {{\bf{f}},{\bf{f'}};{\omega}} \right),
\label{eq:eq_mot_GF}
\end{multline}
where the term
\begin{equation}
\Omega \left( {m'm} \right) = {\varepsilon _{m}} - {\varepsilon _{m'}} - \mu
\label{eq:Omega}
\end{equation}
is the energy of the quasiparticle $\left( {m'm} \right)$. The energy of quasiparticles does not depend on the value of ${\varepsilon _0}$ which means that the latter will not affect the band structure. The quantities $F^{mm'} = \left\langle {{X^{mm}}} \right\rangle + \left\langle {{X^{m'm'}}} \right\rangle $ are the filling factors, where $\left\langle {{X^{mm}}} \right\rangle $ are the filling numbers of the states. The filling of the ground single-hole state components is $\left\langle {{X^{{b_{1 \downarrow }}{b_{1 \downarrow }}}}} \right\rangle = \left\langle {{X^{{b_{1 \uparrow }}{b_{1 \uparrow }}}}} \right\rangle = \frac{1}{2}$, and the zero-hole and two-hole states are empty, $\left\langle {{X^{00}}} \right\rangle = \left\langle {{X^{{L_j}{L_j}}}} \right\rangle = 0$. The Dyson equation in $\mathbf{k}$-space has the form:
\begin{equation}
{\hat D }\left( {{\bf{k}};\omega } \right) = {\left( {\omega  - \hat \Omega  - \hat F{{\hat T}_{\bf{k}}}} \right)^{ - 1}},
\label{eq:Dyson}
\end{equation}
where $\hat \Omega $ and $\hat F$ are the diagonal matrices of quasiparticle energies and their filling factors, respectively, and ${\hat T_{\bf{k}}}$ is the hopping matrix in $k$-space with the elements
\begin{equation}
t^{mm'}_{n'n}(\mathbf{k}) = \sum\limits_{\mathbf{R}} t^{mm'}_{n'n}(\mathbf{R}) e^{\mathrm{i}\mathbf{kR}}.
\end{equation}

\section{Electronic structure without Coulomb interactions}
The band structure of quasiparticle excitations for the CuO monolayer without taking into account Coulomb interactions (Fig.~5a) obtained using the GTB method completely coincides with the band structure obtained as a result of projecting the electron bands within the eight-band $p-d$ model onto the LDA bands~\cite{Slobodchikov2023}. The full width of all bands is $10$ eV. Figure~5b shows the contributions of each of the basis orbitals to the quasiparticle bands. The highest-energy band is formed by the antibonding ${d_{{x^2} - {y^2}}} - \beta $-orbital to which the copper ${d_{{x^2} - {y^2}}}$-orbital makes a larger contribution.

The contribution of the $\lambda $ orbital to the band states is determined by the term $A_\lambda \left(\mathbf{k},\omega\right)$ in the sum of the total spectral density
\begin{multline}
A_\sigma \left(\mathbf{k},\omega\right) = \sum\limits_\lambda A_{\lambda \sigma} \left(\mathbf{k},\omega\right)  = \\
 =  - \frac{1}{\pi} \sum\limits_\lambda \mathrm{Im} G_{\lambda\lambda\sigma} \left(\mathbf{k},\omega + \mathrm{i}\delta \right) = \\
 =  - \frac{1}{\pi} \sum\limits_{\lambda mm'nn'} \gamma_{\lambda\sigma}^{mm'*} \gamma_{\lambda\sigma}^{n'n} \mathrm{\Im} D^{mm'}_{n'n}\left(\mathbf{k},\omega + \mathrm{i}\delta \right),
\label{eq:Akw}
\end{multline}
The contributions of the ${d_{{x^2} - {y^2}}}$- and $\beta $-orbitals to the spectral density of states of the ${d_{{x^2} - {y^2}}} - \beta $ bands and their ratio depend on the wave vector. The Fermi level intersects the band of the antibonding ${d_{{x^2} - {y^2}}} - \beta $-orbital in the middle. Therefore, the multiband $p-d$ model without Coulomb interactions predicts a metallic state for the CuO monolayer.

The band formed predominantly by the ${d_{xz}}$ and ${d_{yz}}$ orbitals of copper lies close to the Fermi level. The states of this band are hybridized to a greater extent with the $\alpha$ molecular oxygen orbital and to a lesser extent with $\gamma$. At the point ${\rm X}=\left( {\pi ,0} \right)$, the state of this band is at a depth of $0.02$ eV from the Fermi level~\cite{Slobodchikov2023}. This band is a part of the set of three bands (Fig.~5b), two of which are widely dispersed due to hybridization with oxygen orbitals and one weakly dispersed which is entirely formed by the ${d_{xz}}$ and ${d_{yz}}$ copper orbitals. The bands with a predominant contribution of ${d_{xz}}$ and ${d_{yz}}$ orbitals extend from the Fermi level at $-4.5$ eV to the low-energy region.

The set of bands formed predominantly by ${d_{3{z^2} - {r^2}}}$- and ${d_{xy}}$-orbitals is in the energy range from $-3$ to $-1$ eV (Fig.~5b). This set contains two weakly dispersed bands with a small admixture of oxygen orbitals. The ratio of the ${d_{3{z^2} - {r^2}}}$- or ${d_{xy}}$-orbital contributions to each of these bands depends on the wave vector. In the lower band, the contribution of the ${d_{3{z^2} - {r^2}}}$ orbital predominates in the nodal $\Gamma$-$M$ ($\Gamma = \left( {0,0} \right)$, $M=\left( {\pi ,\pi } \right)$) and antinodal ${\rm X}$-$M$ directions, while the ${d_{xy}}$ orbital makes a larger contribution in most of the $\Gamma$-${\rm X}$ direction (Fig.~5b). In the upper band, the opposite situation with the predominance of contributions in one or another direction in the Brillouin zone occurs. The states of the bands with a predominant contribution of the ${d_{xy}}$ orbital are hybridized to a greater extent with the oxygen $\gamma $ orbital and to a lesser extent with $\alpha$.

\end{multicols}
\twocolumn

\begin{figure}
\fig[*]{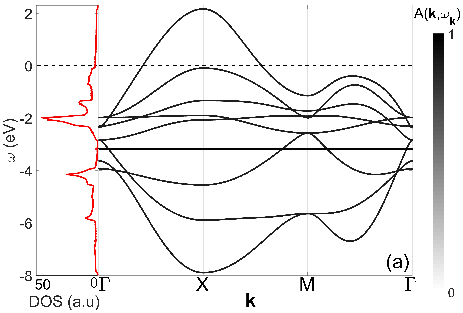}{}{}{*}
\fig[*]{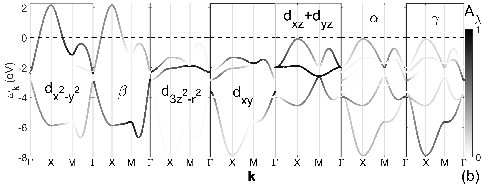}{}{}{*}
\fig[5]{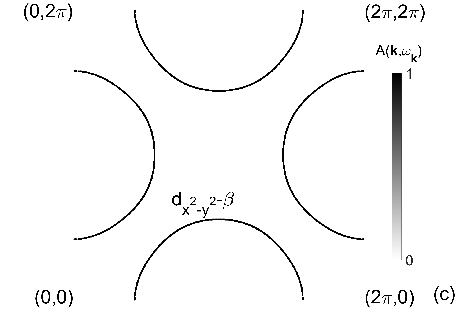}{}{}{(a) Band structure and density of states (DOS) of quasiparticle excitations of the CuO monolayer without Coulomb interactions. (b) Contributions of different orbitals to the bands of quasiparticle excitations. (c) Fermi contour for the CuO monolayer without Coulomb interactions. The following notations are used for the Brillouin zone points: $\Gamma $ --- $\left( {0,0} \right)$, ${\rm X}$ --- $\left( {\pi ,0} \right)$, ${\rm M}$ --- $\left( {\pi ,\pi } \right)$ (in units of $\frac{1}{a}$). The intensity of the black dots shows the total $A\left( {{\bf{k}},{\omega _{\bf{k}}}} \right)$ or partial ${A_\lambda }\left( {{\bf{k}},{\omega _{\bf{k}}}} \right)$ (of the orbital $\lambda$) spectral density of states with the wave vector ${\bf{k}}$ and the energy $\omega_{\bf{k}}$. The correspondence between the color and the magnitude of the spectral density is shown on the gradient scale. From here on, the horizontal dashed line on the band structures at $\omega_{\bf{k}}=0$ shows the chemical potential level $\mu$. The orbitals near the Fermi contour are those that contribute most to the spectral density of its states.}
\end{figure}

It is evident from the distribution of copper and oxygen contributions in the bands and from the bandwidths that the strongest hybridization with oxygen orbitals occurs for the $d_{{x^2} - {y^2}}$- and $d_{xz}$-, $d_{yz}$-orbitals (Fig.~5b). For the ${d_{xy}}$ orbital, the $p-d$ hybridization is weaker, while it is almost absent for the ${d_{3{z^2} - {r^2}}}$ states. Therefore, the nonzero contribution of oxygen orbitals is present only in seven of the eight bands. The oxygen $p$-bands (bands with a predominant contribution of oxygen states) are located in the range from $-8$ to $-3$ eV. Based on the position of the gravity centers of the oxygen bands, we can say that the band with the predominant contribution of the $\alpha$-orbital is located above all, the $\beta$-band is located below it, and the $\gamma$-band is located below all.

The Fermi surface (or more precisely, the Fermi contour since the system under consideration is two-dimensional) of the CuO monolayer without taking into account Coulomb interactions is four hole pockets (Fig.~5c) around each of the points $\left( {\pm\pi ,0} \right)$, $\left( {0,\pm\pi } \right)$ in $\mathbf{k}$-space. These pockets have the nature of copper ${d_{{x^2} - {y^2}}}$- and oxygen ${p_x}$- and ${p_y}$-orbitals. Since the bands with the nature of copper ${d_{xz}}$, ${d_{yz}}$ and oxygen ${p_z}$ orbitals are close to the Fermi level, it is expected that this band can cross the Fermi level under some small perturbation (pressure, doping), and then four more internal pockets around the same $\mathbf{k}$-points can appear on the Fermi contour.

\section{Electronic structure of $CuO$ monolayer considering the Coulomb interactions}
\begin{figure}
\fig[*]{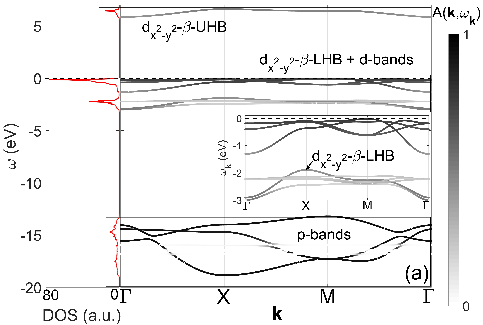}{}{}{*}
\fig[*]{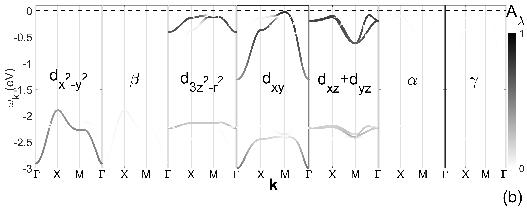}{}{}{*}
\fig[6]{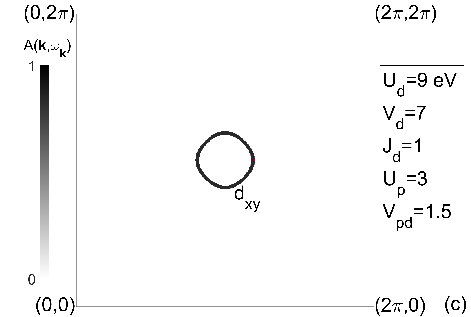}{}{}{Same as in Fig.~5, but at ${U_d}=9$, ${V_d}=7$, ${J_d}=1$, ${U_p}=3$, ${V_{pd}}=1.5$~eV. Panel (c) shows the constant energy map at the top of the valence band (at $-0.025$~eV below the Fermi level).}
\end{figure}
\begin{figure}
\fig[7]{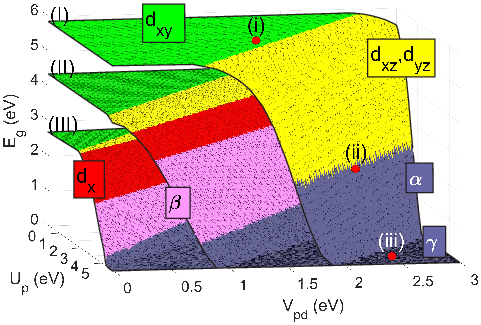}{}{}{Dependencies of the band gap $E_g$ for the CuO monolayer in the paramagnetic phase in the parameter space of the intra-atomic Coulomb interaction on oxygen atoms ${U_p}$ and the interatomic copper-oxygen Coulomb interaction ${V_{pd}}$ for three fixed sets of intra-atomic Coulomb interactions on copper ${U_d}$, ${V_d}$, ${J_d}$: (I) ${U_d}=9$, ${V_d}=7$, ${J_d}=1$; (II) ${U_d}=6$, ${V_d}=5$, ${J_d}=0.5$; (III) ${U_d}=4$, ${V_d}=3.2$, ${J_d}=0.4$~eV. The uniformly colored areas of the surfaces show the regions of the Coulomb parameters for which a certain orbital indicated next to the corresponding area makes a predominant contribution to the states at the top of the valence band. The band gap is indirect at all values of the Coulomb parameters. The points (i), (ii) and (iii) show the values of the parameters ${U_p}$ and ${V_{pd}}$ at which the band structure will be discussed further.}
\end{figure}

The band structure significantly changes when the Coulomb interactions are taken into account. The Coulomb interaction ${U_d}$ leads to splitting of the quasiparticle excitation bands formed by the half-filled copper $d_{{x^2} - {y^2}}$-orbital and the oxygen $\beta$-orbital into the upper (UHB) and lower (LHB) Hubbard subbands ${d_{{x^2} - {y^2}}} - \beta - UHB$ and ${d_{{x^2} - {y^2}}} - \beta - LHB$. The intra-atomic Coulomb interactions on copper increase the energy of electrons in the completely filled bands formed mainly by copper $d$-orbitals. The position of the bands with a predominant contribution of copper orbitals changes relative to the excitations with a predominant contribution of oxygen states depending on the ratio between the parameters of intra-atomic Coulomb interactions on copper atoms ${U_d}$, ${V_d}$, ${J_d}$ and the parameters of similar interactions on oxygen atoms $U_p$ together with ${V_{pd}}$. The change in the relative position of the $d$- and $p$-bands affects the effects of hybridization of copper and oxygen orbitals causing a reconstruction of the dispersion, a redistribution of the spectral weight between the bands, and a redistribution of the contributions of each orbital to the spectral weight of the band states. It is convenient to consider the effects of the influence of the magnitude of various types of Coulomb interactions on the electronic structure as a change in the position and structure of certain blocks of bands. The four blocks of bands can be distinguished (shown in Fig.~6a): the Hubbard subbands ${d_{{x^2} - {y^2}}} - \beta - UHB$ and ${d_{{x^2} - {y^2}}} - \beta - LHB$, the $d$-bands (including the bands of ${d_{3{z^2} - {r^2}}}$, ${d_{xy}}$, ${d_{xz}}$, ${d_{yz}}$-orbitals) and the $p$-bands of the oxygen orbitals $\alpha $, $\beta $ and $\gamma $. The designation of each band using orbitals is based on the principle of which orbitals give a predominant contribution to this band.

At the parameter values $U_p=3$, $V_{pd}=1.5$ (Fig.~7, point (i)), the $d$-bands have a higher energy compared to the $p$-bands. The upper part of the valence band is formed by the $d$-bands, the ${d_{{x^2} - {y^2}}} - \beta - LHB$ subband is in the lower part of the $d$-band group (inset in Fig.~6a). The conductivity band is the upper Hubbard subband ${d_{{x^2} - {y^2}}} - \beta - UHB$. The oxygen $p$-states are located in the depth of the valence band (Fig.~6a) and have almost no effect on the upper part of the valence band, the conductivity band and the magnitude of the band gap. Therefore, the effects of the copper-oxygen hybridization are weakened, and each of the bands acquires a pronounced character of either $d$- or $p$-orbitals (Fig.~6b). A weak effect of the oxygen bands on the band gap value occurs in a wide range of parameters $U_p$ and ${V_{pd}}$ forming a plateau (Fig.~7, surface (I), green color with the label $d_{xy}$) on the surface of the band gap value ${E_g}\left( {{U_p},{V_{pd}}} \right)$ at the fixed parameters ${U_d=9}$, ${V_d=7}$, ${J_d=1}$~eV. Note that this plateau is preserved at other fixed sets of ${U_d}$, ${V_d}$, ${J_d}$ of smaller value, although in a narrower region of the parameters $U_p$, ${V_{pd}}$ (Fig.~7, surfaces (II),(III)). The band gap is open between the subband ${d_{{x^2} - {y^2}}} - \beta - UHB$ and the band of the antibonding orbital $d_{xy}-\alpha-\gamma$ (Fig.~6a,b) with a predominant contribution of the $d_{xy}$ states which forms the top of the valence band (Fig.~7). The maximum of the valence band is at the point ${\rm {M}}$ (Fig.~6b, inset in Fig.~6a), states at a small depth from the top of the valence band create the constant energy map in the form of contour around this point with states of high spectral density (Fig.~6c).

\begin{figure}
\fig[*]{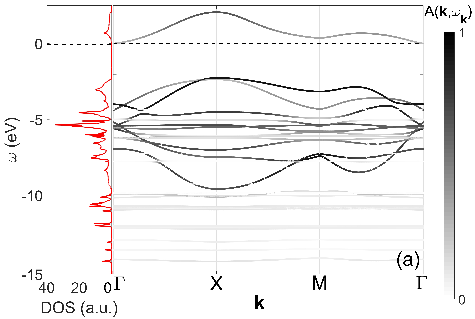}{}{}{*}
\fig[*]{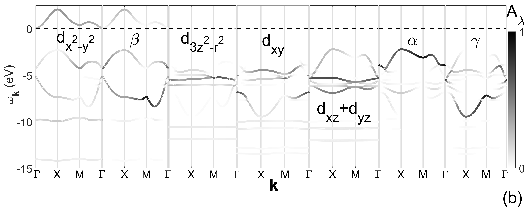}{}{}{*}
\fig[8]{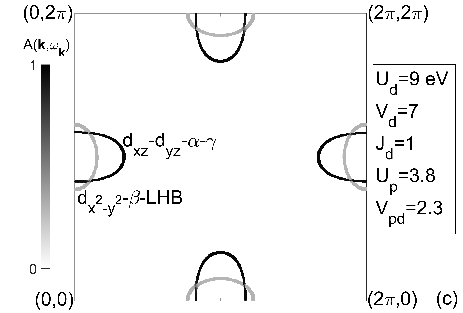}{}{}{Same as in Fig.~5, but at ${U_d}=9$, ${V_d}=7$, ${J_d}=1$, ${U_p}=3.8$, ${V_{pd}}=2.3$~eV. Panel (c) shows the constant energy map at the top of the valence band (at $-2.4$~eV below the Fermi level).}
\end{figure}
\begin{figure}
\fig[*]{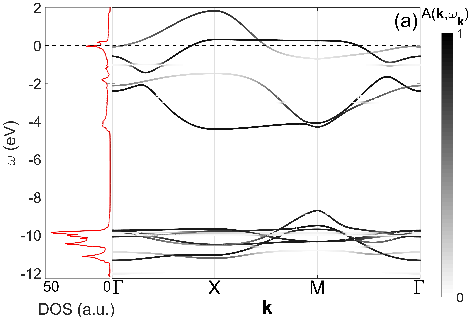}{}{}{*}
\fig[*]{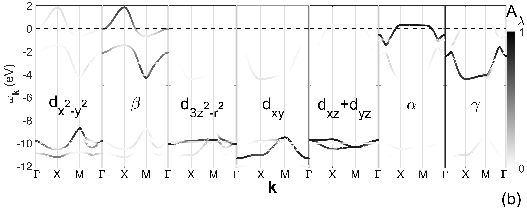}{}{}{*}
\fig[9]{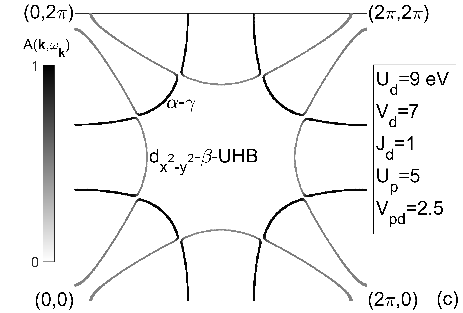}{}{}{Same as in Fig.~5, but at ${U_d}=9$, ${V_d}=7$, ${J_d}=1$, ${U_p}=5$, ${V_{pd}}=2.5$~eV.}
\end{figure}
The intra-atomic interaction on oxygen atoms $U_p$ and interatomic interaction ${V_{pd}}$ increase the energy of electrons in completely filled oxygen orbitals. This is obvious with respect to $U_p$. The number of interatomic interactions of electrons on oxygen orbitals with electrons of neighboring copper atoms (${2\left( {2{l_{Cu}} + 1} \right)}z_{pd}$) is greater than the number of similar interactions of an electron on copper with electrons of neighboring oxygen atoms (${2\left( {2{l_{O}} + 1} \right)}z_{dp}$). This means that the energy of electrons on oxygen will increase compared to the energy of electrons on copper with an increase in $V_{pd}$. The band structure is much more sensitive to the value of the parameter ${V_{pd}}$ than to ${U_p}$.

It is evident from Fig.~7 that when the Coulomb parameters $U_p$ and ${V_{pd}}$ exceed certain values, the nature of the states at the top of the valence band changes and the size of the band gap begins to decrease sharply. The states with the nature of $d_{xz}$-, $d_{yz}$-orbitals emerge at the top of the valence band instead of $d_{xy}$-states. (Fig.~7, yellow color with the label $d_{xz},d_{yz}$). These changes are due to the influence of the $p$-bands: the higher the values of the parameters $U_p$ and ${V_{pd}}$, the higher the energies of the oxygen states, which means that these states begin to influence the $d$-bands located in the upper part of the valence band. The hybridization with oxygen orbitals $\alpha$, $\gamma$ for $d_{xz}$-, $d_{yz}$-orbitals is stronger than for $d_{xy}$-orbital, therefore the contribution of oxygen orbitals to the ${d_{xz}} - {d_{yz}} -\alpha-\gamma$ band will be greater than to the $d_{xy} -\alpha-\gamma$ band. The increase in the width of the antibonding orbital ${d_{xz}} - {d_{yz}} -\alpha-\gamma$ band with an increase in ${U_p}$ and ${V_{pd}}$ occurs faster than for the $d_{xy}-\alpha-\gamma$ band. The appearance of $d_{xz}$-, $d_{yz}$-orbitals at the top of the valence band leads to a decrease in the band gap. The same mechanism leads to an increase in the energies of the states of the ${d_{{x^2} - {y^2}}} - \beta - LHB$ band.

At the parameter values $U_p=3.8$, $V_{pd}=2.3$~eV (Fig.~7, point (ii)), the $p$-bands are already mixed with the $d$-bands (Fig.~8a,b). The top of the valence band is formed by the same ${d_{xz}} - {d_{yz}} -\alpha-\gamma$ band, however the largest contribution to these states is given by the oxygen orbitals (Fig.~8b, Fig.~7, dark blue color with the $\alpha$ label). The ${d_{{x^2} - {y^2}}} - \beta - LHB$ subband is also pushed almost to the top of the valence band (Fig.~8b). The maxima of both ${d_{xz}} - {d_{yz}} -\alpha-\gamma$ and ${d_{{x^2} - {y^2}}} - \beta - LHB$ bands are located at $X$ points (Fig.~8a,b), but the dispersions and spectral densities of these bands differ. The spectral density of the ${d_{xz}} - {d_{yz}} -\alpha-\gamma$ band is close to the maximum. The constant energy map at the top of the valence band includes two types of hole contours around $X$ points: high-intensity ones formed by the ${d_{xz}} - {d_{yz}} -\alpha-\gamma$ band and less intense ones with the states of the ${d_{{x^2} - {y^2}}} - \beta - LHB$ band (Fig.~8c). At parameters ${U_p}=3.8$, ${V_{pd}}=2.3$~eV (Fig.~8a), the oxygen bands are in the same group with the bands of copper orbitals, just as in the case without Coulomb interactions.

At even larger values $U_p=5$ and ${V_{pd}}=2.5$~eV (Fig.~7, point (iii)), the oxygen bands rise above the $d$-bands and enter the conductivity band (Fig.~9a). In this case, the band gap closes (Fig.~7, flat region at $E_g=0$). The Fermi level intersects the ${d_{{x^2} - {y^2}}} - \beta - UHB$ band which is predominantly formed by the oxygen $\beta$-orbital with a small admixture of the $d_{{x^2} - {y^2}}$-orbital and the $\alpha-\gamma$ band which is almost entirely formed by the $\alpha$-orbital with a small contribution of the $\gamma$-orbital (Figs.~9a,b). The $\alpha-\gamma$-band has an almost dispersionless region in the $\mathrm{X}-\mathrm{M}$ and $\mathrm{Y}-\mathrm{M}$ directions. The presence of such a large flat region of the band leads to a high density of states which can be an important factor to achieve high $T_c$ in the superconducting phase of the system with the corresponding Coulomb parameters. The two-dimensional Fermi surface (Fermi contour) for these parameters consists of the four large electron pockets around the points $\Gamma$, $\left( {2\pi ,0} \right)$, $\left( {0,2\pi } \right)$, $\left( {2\pi ,2\pi } \right)$ with the high-intensity $\alpha-\gamma$ states and the four hole pockets around the $\mathrm{X}$ point and the symmetric $-\mathrm{X}$, $\mathrm{Y}$ and $-\mathrm{Y}$ points formed by the ${d_{{x^2} - {y^2}}} - \beta - UHB$ states. However, a splittings occurs at the intersection points of the contours of these two types, and eight contours of complex shape with a non-uniform distribution of spectral weight arise (Fig.~9c).

If we assume that the parameters ${U_d}$, ${V_d}$, ${J_d}$ have smaller values in the CuO monolayer, the general tendency of the transformations of the electronic structure from the insulating to the metallic state with a change in the parameters $U_p$ and ${V_{pd}}$ is preserved: at first, the type of copper orbitals at the top of the valence band changes with an increase in the contribution of oxygen states to the $d$-bands, then the contribution of oxygen states becomes predominant, and then this contribution becomes the only one (Fig.~7, surfaces (I), (II) and (III)). Although the regimes of reduced parameters ${U_d}$, ${V_d}$, ${J_d}$ have their own peculiarities. At ${U_d}=4$, ${V_d}=3.2$, ${J_d}=0.4$ (Fig.~7, surface (III)) and ${U_d}=6$, ${V_d}=5$, ${J_d}=0.5$ (Fig.~7, surface (II)), the region of parameters in which the ${d_{xz}}$, ${d_{yz}}$ states are at the top of the valence band is very narrow. Instead, the states of the ${d_{{x^2} - {y^2}}} - \beta - LHB$ subband with a predominant contribution of $d_{{x^2} - {y^2}}$- (Fig.~7, red color with the $d_{{x^2} - {y^2}}$ label) or $\beta$-orbitals (Fig.~7, pink color with the $\beta$ label) are pushed to the top of the valence band. At even smaller values of intra-atomic Coulomb interactions on copper atoms, the band gap is determined by the splitting between the ${d_{{x^2} - {y^2}}} - \beta - UHB$ and ${d_{{x^2} - {y^2}}} - \beta - LHB$ subbands, the top of the valence band has the nature of $d_{{x^2} - {y^2}}$ orbitals, and the $d_{xy}$, $d_{xz}$, $d_{yz}$ orbitals are always located deep inside the valence band.

\section{Conclusion}

The electronic structure of the CuO monolayer, the magnitude of the band gap, the contributions of various orbitals to the band states, in particular, at the top of the valence band are calculated considering the local Coulomb interactions.

In the regime of strong Coulomb interactions on copper atoms, which prevail over other Coulomb interactions, the electron system is in the insulating phase, the top of the valence band is formed by the copper $d_{xy}$-orbitals in the vicinity of the $M$ point, oxygen states are deep inside the valence band and almost do not contribute to low-energy excitations. The electronic structure depends significantly on the values of the interatomic copper-oxygen Coulomb interaction $V_{pd}$ and the intraatomic Coulomb interaction on oxygen atoms $U_p$. The energy of the oxygen states and their hybridization with $d$-states increases with increasing $V_{pd}$ and $U_p$, as a result of which the dispersion and orbital nature of the states at the top of the valence band are reconstructed, and the size of the band gap decreases. In the process of this reconstruction, first the ${d_{xz}}$, ${d_{yz}}$ states of copper reach the top of the valence band, and then the oxygen orbitals do so. 

The values of the Coulomb parameters $U_p$ and $V_{pd}$ at which the contribution of the oxygen orbitals at the top of the valence band becomes predominant and the ground electronic state of the system becomes metallic are calculated.

The present work was carried out within the state assignment of Kirensky Institute of Physics.

\onecolumn\begin{multicols}{2}
\longformula
\section*{Appendix 1}
The hopping integrals values are presented in tables 2---4.
\begin{table}
\tabl[2]{{\sffamily\small\bfseries } The absolute values of the hopping integrals between the orbitals of the copper atom at the site ${\bf{f}} \equiv \left( {{f_x},{f_y}} \right)$ and the oxygen atom at the site ${\bf{f}} + {\bf{R}} = \left( {{f_x} + {R_x},{f_y} + {R_y}} \right)$ in the eight-band $p-d$ Hamiltonian (in eV)}

\begin{center}
\begin{tabular}{|c|c|c|c|c|c|}
\hline
\rule{0mm}{5mm}
${\bf{R}} = \left( {{R_x},{R_y}} \right)$ & $\left( {{a \mathord{\left/
 {\vphantom {a 2}} \right.
 \kern-\nulldelimiterspace} 2},{b \mathord{\left/
 {\vphantom {b 2}} \right.
 \kern-\nulldelimiterspace} 2}} \right)$ & $\left( {{{ - a} \mathord{\left/
 {\vphantom {{ - a} 2}} \right.
 \kern-\nulldelimiterspace} 2},{b \mathord{\left/
 {\vphantom {b 2}} \right.
 \kern-\nulldelimiterspace} 2}} \right)$ & $\left( { \pm {{3a} \mathord{\left/
 {\vphantom {{3a} 2}} \right.
 \kern-\nulldelimiterspace} 2}, \pm {b \mathord{\left/
 {\vphantom {b 2}} \right.
 \kern-\nulldelimiterspace} 2}} \right)$ & $\left( { \pm {{3a} \mathord{\left/
 {\vphantom {{3a} 2}} \right.
 \kern-\nulldelimiterspace} 2}, \pm {{3b} \mathord{\left/
 {\vphantom {{3b} 2}} \right.
 \kern-\nulldelimiterspace} 2}} \right)$ & $\left( { \pm {{5a} \mathord{\left/
 {\vphantom {{5a} 2}} \right.
 \kern-\nulldelimiterspace} 2}, \pm {b \mathord{\left/
 {\vphantom {b 2}} \right.
 \kern-\nulldelimiterspace} 2}} \right)$ \\
& $\left( {{{ - a} \mathord{\left/
 {\vphantom {{ - a} 2}} \right.
 \kern-\nulldelimiterspace} 2}, - {b \mathord{\left/
 {\vphantom {b 2}} \right.
 \kern-\nulldelimiterspace} 2}} \right)$ & $\left( {{a \mathord{\left/
 {\vphantom {a 2}} \right.
 \kern-\nulldelimiterspace} 2}, - {b \mathord{\left/
 {\vphantom {b 2}} \right.
 \kern-\nulldelimiterspace} 2}} \right)$ & $\left( { \pm {a \mathord{\left/
 {\vphantom {a 2}} \right.
 \kern-\nulldelimiterspace} 2}, \pm {{3b} \mathord{\left/
 {\vphantom {{3b} 2}} \right.
 \kern-\nulldelimiterspace} 2}} \right)$ & & $\left( { \pm {a \mathord{\left/
 {\vphantom {a 2}} \right.
 \kern-\nulldelimiterspace} 2}, \pm {{5b} \mathord{\left/
 {\vphantom {{5b} 2}} \right.
 \kern-\nulldelimiterspace} 2}} \right)$ \\
 \hline
$\left| {{t_{d_{{x^2} - {y^2}}p_x}}\left( {\bf{R}} \right)} \right|$ & $1.349$ & $0$ & $0.04$ & $0.017$ & $0.004$\\
$\left| {{t_{d_{{x^2} - {y^2}}p_y}}\left( {\bf{R}} \right)} \right|$ & $0$ & $1.349$ & $0.04$ & $0$ & $0.004$\\
$\left| {{t_{d_{3{z^2} - {r^2}}p_x}}\left( {\bf{R}} \right)} \right|$ & $0.367$ & $0$ & $0.018$ & $0.003$ & $0.002$\\
$\left| {{t_{d_{3{z^2} - {r^2}}p_y}}\left( {\bf{R}} \right)} \right|$ & $0$ & $0.367$ & $0.018$ & $0$ & $0.002$\\
$\left| {{t_{d_{xy}p_x}}\left( {\bf{R}} \right)} \right|$ & $0$ & $0.782$ & $0.029$ & $0$ & $0.003$\\
$\left| {{t_{d_{xy}p_y}}\left( {\bf{R}} \right)} \right|$ & $0.782$ & $0$ & $0.006$ & $0.01$ & $0.003$\\
$\left| {{t_{d_{xz}p_z}}\left( {\bf{R}} \right)} \right|$ & $0.784$ & $0$ & $0.012$ & $0.01$ & $0.005$\\
$\left| {{t_{d_{yz}p_z}}\left( {\bf{R}} \right)} \right|$ & $0$ & $0.784$ & $0.012$ & $0$ & $0.005$\\
\hline
$\left| {{t_{d_{{x^2} - {y^2}}p_z}}\left( {\bf{R}} \right)} \right|$, $\left| {{t_{d_{3{z^2} - {r^2}}p_z}}\left( {\bf{R}} \right)} \right|$, & & & & &\\
$\left| {{t_{d_{xy}p_z}}\left( {\bf{R}} \right)} \right|$, $\left| {{t_{d_{xz}p_x}}\left( {\bf{R}} \right)} \right|$, & $0$ & $0$ & $0$ & $0$ & $0$\\
 $\left| {{t_{d_{yz}p_x}}\left( {\bf{R}} \right)} \right|$, $\left| {{t_{d_{xz}p_y}}\left( {\bf{R}} \right)} \right|$, & & & & &\\
  $\left| {{t_{d_{yz}p_y}}\left( {\bf{R}} \right)} \right|$ & & & & & \\
\hline
\end{tabular}
\end{center}
\end{table}

\begin{table}
\tabl[3]{{\sffamily\small\bfseries } The absolute values of the hopping integrals between the orbitals of oxygen atoms at the sites ${\bf{g}} \equiv \left( {{g_x},{g_y}} \right)$ and ${\bf{g}} + {\bf{R}} = \left( {{g_x} + {R_x},{g_y} + {R_y}} \right)$ in the eight-band $p-d$ model Hamiltonian (in eV)}

\begin{center}
\begin{tabular}{|c|c|c|c|c|c|}
\hline
\rule{0mm}{5mm}
${\bf{R}} = \left( {{R_x},{R_y}} \right)$ & $\left( { \pm a,0} \right)$ & $\left( { \pm a, \pm b} \right)$ & $\left( { \pm 2a,0} \right)$ & $\left( { \pm 2a, \pm b} \right)$ & $\left( { \pm 2a, \pm 2b} \right)$ \\
& $\left( {0, \pm b} \right)$ & & $\left( {0, \pm 2b} \right)$ & $\left( { \pm a, \pm 2b} \right)$ &\\
 \hline
$\left| {{t_{p_xp_y}}\left( {\bf{R}} \right)} \right|$ & $0.609$ & $0$ & $0.026$ & $0.001$ & $0$\\
$\left| {{t_{p_xp_x}}\left( {\bf{R}} \right)} \right|$, $\left| {{t_{p_yp_y}}\left( {\bf{R}} \right)} \right|$ & $0.353$ & $0.148$ & $0.036$ & $0.008$ & $0.001$\\
$\left| {{t_{p_zp_z}}\left( {\bf{R}} \right)} \right|$ & $0.253$ & $0.069$ & $0.008$ & $0.005$ & $0.002$\\
\hline
\end{tabular}
\end{center}
\end{table}

\begin{table}
\tabl[4]{{\sffamily\small\bfseries } The absolute values of the hopping integrals between the orbitals of copper atoms at the sites ${\bf{f}} \equiv \left( {{f_x},{f_y}} \right)$ and ${\bf{f}} + {\bf{R}} = \left( {{f_x} + {R_x},{f_y} + {R_y}} \right)$ in the eight-band $p-d$ model Hamiltonian (in eV)}

\begin{center}
\begin{tabular}{|c|c|c|c|c|c|}
\hline
\rule{0mm}{5mm}
${\bf{R}} = \left( {{R_x},{R_y}} \right)$ & $\left( { \pm a,0} \right)$ & $\left( { \pm a, \pm b} \right)$ & $\left( { \pm 2a,0} \right)$ & $\left( { \pm 2a, \pm b} \right)$ & $\left( { \pm 2a, \pm 2b} \right)$ \\
& $\left( {0, \pm b} \right)$ & & $\left( {0, \pm 2b} \right)$ & $\left( { \pm a, \pm 2b} \right)$ &\\
 \hline
$\left| {{t_{d_{{x^2} - {y^2}}d_{{x^2} - {y^2}}}}\left( {\bf{R}} \right)} \right|$ & $0.157$ & $0.01$ & $0.017$ & $0.002$ & $0.003$\\
$\left| {{t_{d_{{x^2} - {y^2}}d_{3{z^2} - {r^2}}}}\left( {\bf{R}} \right)} \right|$ & $0$ & $0.003$ & $0$ & $0.003$ & $0$\\
$\left| {{t_{d_{{x^2} - {y^2}}d_{xy}}}\left( {\bf{R}} \right)} \right|$ & $0$ & $0$ & $0$ & $0.003$ & $0$\\
$\left| {{t_{d_{{x^2} - {y^2}}d_{xz}}}\left( {\bf{R}} \right)} \right|$, $\left| {{t_{d_{{x^2} - {y^2}}d_{yz}}}\left( {\bf{R}} \right)} \right|$ & $0$ & $0$ & $0$ & $0$ & $0$\\
$\left| {{t_{d_{3{z^2} - {r^2}}d_{3{z^2} - {r^2}}}}\left( {\bf{R}} \right)} \right|$ & $0.049$ & $0.013$ & $0.02$ & $0.001$ & $0$\\
$\left| {{t_{d_{3{z^2} - {r^2}}d_{xy}}}\left( {\bf{R}} \right)} \right|$ & $0.01$ & $0$ & $0.002$ & $0$ & $0$\\
$\left| {{t_{d_{3{z^2} - {r^2}}d_{xz}}}\left( {\bf{R}} \right)} \right|$, $\left| {{t_{d_{3{z^2} - {r^2}}d_{yz}}}\left( {\bf{R}} \right)} \right|$ & $0$ & $0$ & $0$ & $0$ & $0$\\
$\left| {{t_{d_{xy}d_{xy}}}\left( {\bf{R}} \right)} \right|$ & $0.243$ & $0.04$ & $0.003$ & $0.002$ & $0.001$\\
$\left| {{t_{d_{xy}d_{xz}}}\left( {\bf{R}} \right)} \right|$, $\left| {{t_{d_{xy}d_{yz}}}\left( {\bf{R}} \right)} \right|$ & $0$ & $0$ & $0$ & $0$ & $0$\\
$\left| {{t_{d_{xz}d_{xz}}}\left( {\bf{R}} \right)} \right|$, $\left| {{t_{d_{yz}d_{yz}}}\left( {\bf{R}} \right)} \right|$ & $0.074$ & $0.051$ & $0.02$ & $0$ & $0$\\
$\left| {{t_{d_{xz}d_{yz}}}\left( {\bf{R}} \right)} \right|$ & $0.04$ & $0$ & $0.003$ & $0.001$ & $0$\\
\hline
\end{tabular}
\end{center}
\end{table}

\section*{Appendix 2}
The Shastry-type orthogonalization procedure for three orbitals is the transformation from the atomic orbitals ${p_{x{\bf{k}}}}$, ${p_{y{\bf{k}}}}$, ${p_{z{\bf{k}}}}$ to new molecular orbitals ${\alpha _{\bf{k}}}$, ${\beta _{\bf{k}}}$, ${\gamma _{\bf{k}}}$:
\begin{equation}
\left[ {\begin{array}{*{20}{c}}
{{\alpha _{\bf{k}}}}\\
{{\beta _{\bf{k}}}}\\
{{\gamma _{\bf{k}}}}
\end{array}} \right] = \left[ {\begin{array}{*{20}{c}}
{\frac{{{f_{2{\bf{k}}}}{f_{3{\bf{k}}}}}}{{{\mu _{\bf{k}}}{\eta _{\bf{k}}}}}}&{\frac{{{f_{1{\bf{k}}}}{f_{3{\bf{k}}}}}}{{{\mu _{\bf{k}}}{\eta _{\bf{k}}}}}}&{\frac{{{\mu _{\bf{k}}}}}{{{\eta _{\bf{k}}}}}}\\
{\frac{{i{f_{1{\bf{k}}}}}}{{{\mu _{\bf{k}}}}}}&{ - \frac{{i{f_{2{\bf{k}}}}}}{{{\mu _{\bf{k}}}}}}&0\\
{ - \frac{{i{f_{2{\bf{k}}}}}}{{{\eta _{\bf{k}}}}}}&{ - \frac{{i{f_{1{\bf{k}}}}}}{{{\eta _{\bf{k}}}}}}&{\frac{{i{f_{3{\bf{k}}}}}}{{{\eta _{\bf{k}}}}}}
\end{array}} \right]\left[ {\begin{array}{*{20}{c}}
{{p_{x{\bf{k}}}}}\\
{{p_{y{\bf{k}}}}}\\
{{p_{z{\bf{k}}}}}
\end{array}} \right]\,
 \label{eq:Shastry}
\end{equation}
where ${\mu _{\bf{k}}} = \sqrt {f_{1{\bf{k}}}^2 + f_{2{\bf{k}}}^2} $, ${\eta _{\bf{k}}} = \sqrt {f_{1{\bf{k}}}^2 + f_{2{\bf{k}}}^2 + f_{3{\bf{k}}}^2} $, and the real functions ${f_{1{\bf{k}}}}$, ${f_{2{\bf{k}}}}$, ${f_{3{\bf{k}}}}$ are chosen as follows:
\begin{multline}
{f_{1{\bf{k}}}} = \sin \left( {\frac{{{k_x} + {k_y}}}{2}} \right)\\
{f_{2{\bf{k}}}} = \sin \left( {\frac{{{k_x} - {k_y}}}{2}} \right)\\
{f_{3{\bf{k}}}} = \frac{1}{2}\left( {\sin \left( {\frac{{{k_x} + {k_y}}}{2}} \right) + \sin \left( {\frac{{{k_x} - {k_y}}}{2}} \right)} \right)
\label{eq:f_functions}
\end{multline}
The standard Fermi commutation relations remain valid for the operators corresponding to the new orbitals ${\alpha _{\bf{f}}}$, ${\beta _{\bf{f}}}$, ${\gamma _{\bf{f}}}$

\section*{Appendix 3}
The energy of the vacuum hole state ${\varepsilon _0}$ is determined by the expression:
%\longformula
\begin{multline}
{\varepsilon _0} = \sum\limits_{\bf{f}} {\left( {E_{0d{\bf{f}}}^{\left( h \right)} - 10\mu } \right)}  + \sum\limits_{\bf{g}} {\left( {E_{0p{\bf{g}}}^{\left( h \right)} - 6\mu } \right)}  + \sum\limits_{\bf{f}} {E_{0pd{\bf{f}}}^{\left( h \right)}}, \\
E_{0d{\bf{f}}}^{\left( h \right)} = 2\sum\limits_\zeta  {{\varepsilon _{d\zeta }}}  + \left( {2{l_{Cu}} + 1} \right){U_d} +\\
4\left( {2{l_{Cu}} + 1} \right){l_{Cu}}\left( {{V_d} + {J_d}} \right), \\
{d_\zeta } = {d_{{x^2} - {y^2}}},{d_{3{z^2} - {r^2}}},{d_{xy}},{d_{xz}},{d_{yz}},\\
E_{0p{\bf{g}}}^{\left( h \right)} = 2\sum\limits_\xi  {{\varepsilon _{p\xi }}}  + 5\left( {2{l_O} + 1} \right){l_O}{U_p},\,\,{p_\xi } = {p_x},{p_y},{p_z}\\
E_{0pd{\bf{f}}}^{\left( h \right)} = 4{z_{pd}}\left( {2{l_{Cu}} + 1} \right)\left( {2{l_O} + 1} \right){V_{pd}}
\label{eq:E0}
\end{multline}
%\endlongformula

The energy $E_{0d{\bf{f}}}^{\left( h \right)}$ includes the on-site energies of two electrons in each ${d_\zeta }$-orbital of the copper atom at the ${\bf{f}}$ sites, the intra-orbital Coulomb interactions in the $2{l_{Cu}} + 1$ copper orbitals, the inter-orbital Coulomb and the Hund interactions with electrons in the remaining $2{l_{Cu}}$ ${d_\zeta }$-orbitals. $E_{0p{\bf{g}}}^{\left( h \right)}$ is the analogous sum of energies for the electrons on the oxygen atoms at the ${\bf{g}}$ sites with completely filled $2{l_O} + 1$ ${p_{\xi}}$ orbitals, the intra- and inter-orbital Coulomb interactions are considered equal in magnitude (${U_p}$). $E_{0pd{\bf{f}}}^{\left( h \right)}$ contains the Coulomb interactions between the ten copper electrons at the ${\bf{f}}$ sites and the six oxygen electrons on each of the four nearest oxygen atoms. The energy ${\varepsilon _0}$ is the reference level for the energies of the multiparticle cluster eigenstates, but it is not constant (like, for example, the vacuum electron state $\left| 0 \right\rangle $ with zero energy), since ${\varepsilon _0}$ depends on the magnitudes of the Coulomb interactions.

\section*{Appendix 4}
The energies of the oxygen molecular orbitals and the renormalized hopping integrals in the Hamiltonian (\ref{eq:Ham_mo}) are determined by the expressions:

\begin{multline}
\nu _{{\bf{fg}}}^{\left( i \right)} = \frac{1}{N}\sum\limits_{\bf{k}} {\nu _{\bf{k}}^{\left( i \right)}} {e^{ - i{\bf{k}}\left( {{\bf{f}} - {\bf{g}}} \right)}},\,\,i = 1,2,3;\,\,\,\,\,\,\,\kappa _{\bf{R}}^{\left( {\lambda {\rho _j}} \right)} = \frac{1}{N}\sum\limits_{\bf{k}} {\kappa _{\bf{k}}^{\left( {\lambda {\rho _j}} \right)}{e^{i{\bf{kR}}}}} ,\,\,j = 1,2,3,\\
\nu _{{\bf{fg}}}^{\left( 1 \right)} = \frac{{\left( {{\varepsilon _p}f_{3{\bf{k}}}^2 + {\varepsilon _{p_z}}\mu _{\bf{k}}^2} \right)}}{{\eta _{\bf{k}}^2}} + \left( {\frac{{f_{2{\bf{k}}}^2f_{3{\bf{k}}}^2}}{{\eta _{\bf{k}}^2\mu _{\bf{k}}^2}}t_{\bf{k}}^{p_xp_x} + \frac{{f_{1{\bf{k}}}^2f_{3{\bf{k}}}^2}}{{\eta _{\bf{k}}^2\mu _{\bf{k}}^2}}t_{\bf{k}}^{p_yp_y} + \frac{{\mu _{\bf{k}}^2}}{{\eta _{\bf{k}}^2}}t_{\bf{k}}^{p_zp_z}} \right) + \frac{{{f_{1{\bf{k}}}}{f_{2{\bf{k}}}}f_{3{\bf{k}}}^2}}{{\eta _{\bf{k}}^2\mu _{\bf{k}}^2}}\left[ {t_{\bf{k}}^{p_xp_y} + t_{\bf{k}}^{p_yp_x}} \right] + \\
 + \frac{{{f_{2{\bf{k}}}}{f_{3{\bf{k}}}}}}{{\eta _{\bf{k}}^2}}\left[ {t_{\bf{k}}^{p_xp_z} + t_{\bf{k}}^{p_zp_x}} \right] + \frac{{{f_{1{\bf{k}}}}{f_{3{\bf{k}}}}}}{{\eta _{\bf{k}}^2}}\left[ {t_{\bf{k}}^{p_yp_z} + t_{\bf{k}}^{p_zp_y}} \right],\\
\nu _{\bf{k}}^{\left( 2 \right)} = {\varepsilon _p} + \left( {\frac{{f_{1{\bf{k}}}^2}}{{\mu _{\bf{k}}^2}}t_{\bf{k}}^{p_xp_x} + \frac{{f_{2{\bf{k}}}^2}}{{\mu _{\bf{k}}^2}}t_{\bf{k}}^{p_yp_y} - \frac{{{f_{1{\bf{k}}}}{f_{2{\bf{k}}}}}}{{\mu _{\bf{k}}^2}}\left[ {t_{\bf{k}}^{p_xp_y} + t_{\bf{k}}^{p_yp_x}} \right]} \right),\\
\nu _{\bf{k}}^{\left( 3 \right)} = \frac{{\left( {{\varepsilon _p}\mu _{\bf{k}}^2 + {\varepsilon _{p_z}}f_{3{\bf{k}}}^2} \right)}}{{\eta _{\bf{k}}^2}} - \left( {\frac{{f_{2{\bf{k}}}^2}}{{\eta _{\bf{k}}^2}}t_{\bf{k}}^{p_xp_x} + \frac{{f_{1{\bf{k}}}^2}}{{\eta _{\bf{k}}^2}}t_{\bf{k}}^{p_yp_y} + \frac{{f_{3{\bf{k}}}^2}}{{\eta _{\bf{k}}^2}}t_{\bf{k}}^{p_zp_z}} \right) - \frac{{{f_{1{\bf{k}}}}{f_{2{\bf{k}}}}}}{{\eta _{\bf{k}}^2}}\left[ {t_{\bf{k}}^{p_xp_y} + t_{\bf{k}}^{p_yp_x}} \right]\\
 + \frac{{{f_{2{\bf{k}}}}{f_{3{\bf{k}}}}}}{{\eta _{\bf{k}}^2}}\left[ {t_{\bf{k}}^{p_xp_z} + t_{\bf{k}}^{p_zp_x}} \right] + \frac{{{f_{1{\bf{k}}}}{f_{3{\bf{k}}}}}}{{\eta _{\bf{k}}^2}}\left[ {t_{\bf{k}}^{p_yp_z} + t_{\bf{k}}^{p_zp_y}} \right],\\
\nu _{\bf{k}}^{\left( {12} \right)} =  - \frac{{i{f_{1{\bf{k}}}}{f_{2{\bf{k}}}}{f_{3{\bf{k}}}}}}{{{\eta _{\bf{k}}}\mu _{\bf{k}}^2}}t_{\bf{k}}^{p_xp_x} + \frac{{i{f_{1{\bf{k}}}}{f_{2{\bf{k}}}}{f_{3{\bf{k}}}}}}{{{\eta _{\bf{k}}}\mu _{\bf{k}}^2}}t_{\bf{k}}^{p_yp_y} + \frac{{if_{2{\bf{k}}}^2{f_{3{\bf{k}}}}}}{{{\eta _{\bf{k}}}\mu _{\bf{k}}^2}}t_{\bf{k}}^{p_xp_y} - \frac{{if_{1{\bf{k}}}^2{f_{3{\bf{k}}}}}}{{{\eta _{\bf{k}}}\mu _{\bf{k}}^2}}t_{\bf{k}}^{p_yp_x} - \frac{{i{f_{1{\bf{k}}}}}}{{{\eta _{\bf{k}}}}}t_{\bf{k}}^{p_zp_x} + \frac{{i{f_{2{\bf{k}}}}}}{{{\eta _{\bf{k}}}}}t_{\bf{k}}^{p_zp_y},\\
\nu _{\bf{k}}^{\left( {13} \right)} = \frac{{if_{2{\bf{k}}}^2{f_{3{\bf{k}}}}}}{{\eta _{\bf{k}}^2{\mu _{\bf{k}}}}}t_{\bf{k}}^{p_xp_x} + \frac{{if_{1{\bf{k}}}^2{f_{3{\bf{k}}}}}}{{\eta _{\bf{k}}^2{\mu _{\bf{k}}}}}t_{\bf{k}}^{p_yp_y} - \frac{{i{\mu _{\bf{k}}}{f_{3{\bf{k}}}}}}{{\eta _{\bf{k}}^2}}t_{\bf{k}}^{p_zp_z} + \frac{{i{f_{1{\bf{k}}}}{f_{2{\bf{k}}}}{f_{3{\bf{k}}}}}}{{\eta _{\bf{k}}^2{\mu _{\bf{k}}}}}\left[ {t_{\bf{k}}^{p_xp_y} + t_{\bf{k}}^{p_yp_x}} \right] - \frac{{i{f_{2{\bf{k}}}}f_{3{\bf{k}}}^2}}{{\eta _{\bf{k}}^2{\mu _{\bf{k}}}}}t_{\bf{k}}^{p_xp_z} + \\
 + \frac{{i{f_{2{\bf{k}}}}{\mu _{\bf{k}}}}}{{\eta _{\bf{k}}^2}}t_{\bf{k}}^{p_zp_x} + \frac{{i{f_{1{\bf{k}}}}{\mu _{\bf{k}}}}}{{\eta _{\bf{k}}^2}}t_{\bf{k}}^{p_zp_y} - \frac{{i{f_{1{\bf{k}}}}f_{3{\bf{k}}}^2}}{{\eta _{\bf{k}}^2{\mu _{\bf{k}}}}}t_{\bf{k}}^{p_yp_z},\\
\nu _{\bf{k}}^{\left( {23} \right)} =  - \frac{{{f_{1{\bf{k}}}}{f_{2{\bf{k}}}}}}{{{\eta _{\bf{k}}}{\mu _{\bf{k}}}}}t_{\bf{k}}^{p_xp_x} + \frac{{{f_{1{\bf{k}}}}{f_{2{\bf{k}}}}}}{{{\eta _{\bf{k}}}{\mu _{\bf{k}}}}}t_{\bf{k}}^{p_yp_y} + \frac{{f_{2{\bf{k}}}^2}}{{{\eta _{\bf{k}}}{\mu _{\bf{k}}}}}t_{\bf{k}}^{p_yp_x} - \frac{{f_{1{\bf{k}}}^2}}{{{\eta _{\bf{k}}}{\mu _{\bf{k}}}}}t_{\bf{k}}^{p_xp_y} + \frac{{{f_{1{\bf{k}}}}{f_{3{\bf{k}}}}}}{{{\eta _{\bf{k}}}{\mu _{\bf{k}}}}}t_{\bf{k}}^{p_xp_z} - \frac{{{f_{2{\bf{k}}}}{f_{3{\bf{k}}}}}}{{{\eta _{\bf{k}}}{\mu _{\bf{k}}}}}t_{\bf{k}}^{p_yp_z},\\
\kappa _{\bf{k}}^{\left( {d_{{x^2} - {y^2}}\alpha } \right)} = \frac{{{f_{2{\bf{k}}}}{f_{3{\bf{k}}}}}}{{{\mu _{\bf{k}}}{\eta _{\bf{k}}}}}t_{\bf{k}}^{d_{{x^2} - {y^2}}p_x} + \frac{{{f_{1{\bf{k}}}}{f_{3{\bf{k}}}}}}{{{\eta _{\bf{k}}}{\mu _{\bf{k}}}}}t_{\bf{k}}^{d_{{x^2} - {y^2}}p_y} + \frac{{{\mu _{\bf{k}}}}}{{{\eta _{\bf{k}}}}}t_{\bf{k}}^{d_{{x^2} - {y^2}}p_z},\\
\kappa _{\bf{k}}^{\left( {d_{{x^2} - {y^2}}\beta } \right)} =  - \frac{{i{f_{1{\bf{k}}}}}}{{{\mu _{\bf{k}}}}}t_{\bf{k}}^{d_{{x^2} - {y^2}}p_x} + \frac{{i{f_{2{\bf{k}}}}}}{{{\mu _{\bf{k}}}}}t_{\bf{k}}^{d_{{x^2} - {y^2}}p_y},\\
\kappa _{\bf{k}}^{\left( {d_{{x^2} - {y^2}}\gamma } \right)} = \frac{{i{f_{2{\bf{k}}}}}}{{{\eta _{\bf{k}}}}}t_{\bf{k}}^{d_{{x^2} - {y^2}}p_x} + \frac{{i{f_{1{\bf{k}}}}}}{{{\eta _{\bf{k}}}}}t_{\bf{k}}^{d_{{x^2} - {y^2}}p_y} - \frac{{i{f_{3{\bf{k}}}}}}{{{\eta _{\bf{k}}}}}t_{\bf{k}}^{d_{{x^2} - {y^2}}p_z},\,\,\\
\kappa _{\bf{k}}^{\left( {d_{3{z^2} - {r^2}}\alpha } \right)} = \frac{{{f_{2{\bf{k}}}}{f_{3{\bf{k}}}}}}{{{\mu _{\bf{k}}}{\eta _{\bf{k}}}}}t_{\bf{k}}^{d_{3{z^2} - {r^2}}p_x} + \frac{{{f_{1{\bf{k}}}}{f_{3{\bf{k}}}}}}{{{\eta _{\bf{k}}}{\mu _{\bf{k}}}}}t_{\bf{k}}^{d_{3{z^2} - {r^2}}p_y} + \frac{{{\mu _{\bf{k}}}}}{{{\eta _{\bf{k}}}}}t_{\bf{k}}^{d_{3{z^2} - {r^2}}p_z},\\
\kappa _{\bf{k}}^{\left( {d_{3{z^2} - {r^2}}\beta } \right)} =  - \frac{{i{f_{1{\bf{k}}}}}}{{{\mu _{\bf{k}}}}}t_{\bf{k}}^{d_{3{z^2} - {r^2}}p_x} + \frac{{i{f_{2{\bf{k}}}}}}{{{\mu _{\bf{k}}}}}t_{\bf{k}}^{d_{3{z^2} - {r^2}}p_y},\,\\
\kappa _{\bf{k}}^{\left( {d_{3{z^2} - {r^2}}\gamma } \right)} = \frac{{i{f_{2{\bf{k}}}}}}{{{\eta _{\bf{k}}}}}t_{\bf{k}}^{d_{3{z^2} - {r^2}}p_x} + \frac{{i{f_{1{\bf{k}}}}}}{{{\eta _{\bf{k}}}}}t_{\bf{k}}^{d_{3{z^2} - {r^2}}p_y} - \frac{{i{f_{3{\bf{k}}}}}}{{{\eta _{\bf{k}}}}}t_{\bf{k}}^{d_{3{z^2} - {r^2}}p_z}\\
\kappa _{\bf{k}}^{\left( {d_{xy}\alpha } \right)} = \frac{{{f_{2{\bf{k}}}}{f_{3{\bf{k}}}}}}{{{\mu _{\bf{k}}}{\eta _{\bf{k}}}}}t_{\bf{k}}^{d_{xy}p_x} + \frac{{{f_{1{\bf{k}}}}{f_{3{\bf{k}}}}}}{{{\eta _{\bf{k}}}{\mu _{\bf{k}}}}}t_{\bf{k}}^{d_{xy}p_y} + \frac{{{\mu _{\bf{k}}}}}{{{\eta _{\bf{k}}}}}t_{\bf{k}}^{d_{xy}p_z},\,\kappa _{\bf{k}}^{\left( {d_{xy}\beta } \right)} =  - \frac{{i{f_{1{\bf{k}}}}}}{{{\mu _{\bf{k}}}}}t_{\bf{k}}^{d_{xy}p_x} + \frac{{i{f_{2{\bf{k}}}}}}{{{\mu _{\bf{k}}}}}t_{\bf{k}}^{d_{xy}p_y},\\
\kappa _{\bf{k}}^{\left( {d_{xy}\gamma } \right)} = \frac{{i{f_{2{\bf{k}}}}}}{{{\eta _{\bf{k}}}}}t_{\bf{k}}^{d_{xy}p_x} + \frac{{i{f_{1{\bf{k}}}}}}{{{\eta _{\bf{k}}}}}t_{\bf{k}}^{d_{xy}p_y} - \frac{{i{f_{3{\bf{k}}}}}}{{{\eta _{\bf{k}}}}}t_{\bf{k}}^{d_{xy}p_z},\,\\
\kappa _{\bf{k}}^{\left( {d_{xz}\alpha } \right)} = \frac{{{f_{2{\bf{k}}}}{f_{3{\bf{k}}}}}}{{{\mu _{\bf{k}}}{\eta _{\bf{k}}}}}t_{\bf{k}}^{d_{xz}p_x} + \frac{{{f_{1{\bf{k}}}}{f_{3{\bf{k}}}}}}{{{\eta _{\bf{k}}}{\mu _{\bf{k}}}}}t_{\bf{k}}^{d_{xz}p_y} + \frac{{{\mu _{\bf{k}}}}}{{{\eta _{\bf{k}}}}}t_{\bf{k}}^{d_{xz}p_z},\,\kappa _{\bf{k}}^{\left( {d_{xz}\beta } \right)} =  - \frac{{i{f_{1{\bf{k}}}}}}{{{\mu _{\bf{k}}}}}t_{\bf{k}}^{d_{xz}p_x} + \frac{{i{f_{2{\bf{k}}}}}}{{{\mu _{\bf{k}}}}}t_{\bf{k}}^{d_{xz}p_y},\,\\
\kappa _{\bf{k}}^{\left( {d_{xz}\gamma } \right)} = \frac{{i{f_{2{\bf{k}}}}}}{{{\eta _{\bf{k}}}}}t_{\bf{k}}^{d_{xz}p_x} + \frac{{i{f_{1{\bf{k}}}}}}{{{\eta _{\bf{k}}}}}t_{\bf{k}}^{d_{xz}p_y} - \frac{{i{f_{3{\bf{k}}}}}}{{{\eta _{\bf{k}}}}}t_{\bf{k}}^{d_{xz}p_z},\,\,\,\\
\kappa _{\bf{k}}^{\left( {d_{yz}\alpha } \right)} = \frac{{{f_{2{\bf{k}}}}{f_{3{\bf{k}}}}}}{{{\mu _{\bf{k}}}{\eta _{\bf{k}}}}}t_{\bf{k}}^{d_{yz}p_x} + \frac{{{f_{1{\bf{k}}}}{f_{3{\bf{k}}}}}}{{{\eta _{\bf{k}}}{\mu _{\bf{k}}}}}t_{\bf{k}}^{d_{yz}p_y} + \frac{{{\mu _{\bf{k}}}}}{{{\eta _{\bf{k}}}}}t_{\bf{k}}^{d_{yz}p_z},\kappa _{\bf{k}}^{\left( {d_{yz}\beta } \right)} =  - \frac{{i{f_{1{\bf{k}}}}}}{{{\mu _{\bf{k}}}}}t_{\bf{k}}^{d_{yz}p_x} + \frac{{i{f_{2{\bf{k}}}}}}{{{\mu _{\bf{k}}}}}t_{\bf{k}}^{d_{yz}p_y},\\
\kappa _{\bf{k}}^{\left( {d_{yz}\gamma } \right)} = \frac{{i{f_{2{\bf{k}}}}}}{{{\eta _{\bf{k}}}}}t_{\bf{k}}^{d_{yz}p_x} + \frac{{i{f_{1{\bf{k}}}}}}{{{\eta _{\bf{k}}}}}t_{\bf{k}}^{d_{yz}p_y} - \frac{{i{f_{3{\bf{k}}}}}}{{{\eta _{\bf{k}}}}}t_{\bf{k}}^{d_{yz}p_z},\,\,\,
\label{eq:coef_Hp}
\end{multline}
\endlongformula
where $t_{\bf{k}}^{{\lambda}{\lambda '}}$ is the Fourier transform of the hopping integral ${t_{{\lambda}{\lambda '}}}\left( {\bf{R}} \right)$.

The coefficients $\Psi _{{\bf{ff'gg'}}}^{{\rho _i}{\rho _{i'}}{\rho _j}{\rho _{j'}}} $ and $\Phi _{{\bf{fgg'}}}^{{\rho _i}{\rho _j}}$ in the Hamiltonian (\ref{eq:Ham_mo}) are expressed as follows:
\longformula
\begin{multline}
\Psi _{{\bf{ff'gg'}}}^{{\rho _i}{\rho _{i'}}{\rho _j}{\rho _{j'}}} =  \frac{1}{{{N^3}}}\sum\limits_{\zeta \zeta '} {\sum\limits_{{\bf{kqm}}} {S_{i\zeta {\bf{k}}}^*{S_{i'\zeta {\bf{q}}}}S_{j\zeta '{\bf{m}}}^*{S_{j'\zeta '\left( {{\bf{k}} - {\bf{q}} + {\bf{m}}} \right)}}{e^{ - i{\bf{k}}\left( {{\bf{f}} - {\bf{g'}}} \right)}}{e^{i{\bf{q}}\left( {{\bf{f'}} - {\bf{g'}}} \right)}}{e^{ - i{\bf{m}}\left( {{\bf{g}} - {\bf{g'}}} \right)}}} } \\
{S_{1{p_x}{\bf{k}}}} \equiv {S_{\alpha {p_x}{\bf{k}}}} = \frac{{{f_{2{\bf{k}}}}{f_{3{\bf{k}}}}}}{{{\eta _{\bf{k}}}{\mu _{\bf{k}}}}},\,\,\,{S_{2{p_x}{\bf{k}}}} \equiv {S_{\beta {p_x}{\bf{k}}}} =  - \frac{{i{f_{1{\bf{k}}}}}}{{{\mu _{\bf{k}}}}},\,\,\,{S_{3{p_x}{\bf{k}}}} \equiv {S_{\gamma {p_x}{\bf{k}}}} = \frac{{i{f_{2{\bf{k}}}}}}{{{\eta _{\bf{k}}}}}\\
{S_{1{p_y}{\bf{k}}}} = \frac{{{f_{1{\bf{k}}}}{f_{3{\bf{k}}}}}}{{{\eta _{\bf{k}}}{\mu _{\bf{k}}}}},\,\,{S_{2{p_y}{\bf{k}}}} = \frac{{i{f_{2{\bf{k}}}}}}{{{\mu _{\bf{k}}}}},\,\,\,{S_{3{p_y}{\bf{k}}}} = \frac{{i{f_{1{\bf{k}}}}}}{{{\eta _{\bf{k}}}}}\,\\
{S_{1{p_z}{\bf{k}}}} = \frac{{{\mu _{\bf{k}}}}}{{{\eta _{\bf{k}}}}},\,\,\,\,\,\,{S_{2{p_z}{\bf{k}}}} = 0,\,\,\,\,\,\,{S_{3{p_z}{\bf{k}}}} =  - \frac{{i{f_{3{\bf{k}}}}}}{{{\eta _{\bf{k}}}}}\,\,\\
\Phi _{{\bf{fgg'}}}^{{\rho _i}{\rho _j}} = \frac{1}{{{N^2}}}\sum\limits_{{\bf{kq}}} {\Phi _{{\bf{kq}}}^{{\rho _i}{\rho _j}}{C_{{\bf{kq}}}}\left( {{\bf{f}},{\bf{g}},{\bf{g'}}} \right)} ,\\
{C_{{\bf{kq}}}}\left( {{\bf{f}},{\bf{g}},{\bf{g'}}} \right) = 2\left[ {\cos \left( {\left( {{\bf{k}} - {\bf{q}}} \right)\left( {{a \mathord{\left/
 {\vphantom {a {2 + {b \mathord{\left/
 {\vphantom {b 2}} \right.
 \kern-\nulldelimiterspace} 2}}}} \right.
 \kern-\nulldelimiterspace} {2 + {b \mathord{\left/
 {\vphantom {b 2}} \right.
 \kern-\nulldelimiterspace} 2}}}} \right)} \right) + \cos \left( {\left( {{\bf{k}} - {\bf{q}}} \right)\left( {{a \mathord{\left/
 {\vphantom {a {2 - {b \mathord{\left/
 {\vphantom {b 2}} \right.
 \kern-\nulldelimiterspace} 2}}}} \right.
 \kern-\nulldelimiterspace} {2 - {b \mathord{\left/
 {\vphantom {b 2}} \right.
 \kern-\nulldelimiterspace} 2}}}} \right)} \right)} \right]{e^{ - i{\bf{k}}\left( {{\bf{g}} - {\bf{f}}} \right)}}{e^{i{\bf{q}}\left( {{\bf{g'}} - {\bf{f}}} \right)}},\\
\Phi _{{\bf{kq}}}^{\alpha \alpha } = \frac{{{f_{2{\bf{k}}}}{f_{3{\bf{k}}}}}}{{{\eta _{\bf{k}}}{\mu _{\bf{k}}}}}\frac{{{f_{2{\bf{n}}}}{f_{3{\bf{n}}}}}}{{{\eta _{\bf{n}}}{\mu _{\bf{n}}}}} + \frac{{{f_{1{\bf{k}}}}{f_{3{\bf{k}}}}}}{{{\eta _{\bf{k}}}{\mu _{\bf{k}}}}}\frac{{{f_{1{\bf{n}}}}{f_{3{\bf{n}}}}}}{{{\eta _{\bf{n}}}{\mu _{\bf{n}}}}} + \frac{{{\mu _{\bf{k}}}}}{{{\eta _{\bf{k}}}}}\frac{{{\mu _{\bf{n}}}}}{{{\eta _{\bf{n}}}}},\,\,\,\Phi _{{\bf{kq}}}^{\beta \beta } = \left[ {\frac{{{f_{1{\bf{k}}}}}}{{{\mu _{\bf{k}}}}}\frac{{{f_{1{\bf{q}}}}}}{{{\mu _{\bf{q}}}}} + \frac{{{f_{2{\bf{k}}}}}}{{{\mu _{\bf{k}}}}}\frac{{{f_{2{\bf{q}}}}}}{{{\mu _{\bf{q}}}}}} \right],\\
\Phi _{{\bf{kq}}}^{\gamma \gamma } = \left[ {\frac{{{f_{2{\bf{k}}}}}}{{{\eta _{\bf{k}}}}}\frac{{{f_{2{\bf{q}}}}}}{{{\eta _{\bf{q}}}}} + \frac{{{f_{1{\bf{k}}}}}}{{{\eta _{\bf{k}}}}}\frac{{{f_{1{\bf{q}}}}}}{{{\eta _{\bf{q}}}}} + \frac{{{f_{3{\bf{k}}}}}}{{{\eta _{\bf{k}}}}}\frac{{{f_{3{\bf{q}}}}}}{{{\eta _{\bf{q}}}}}} \right],\\
\Phi _{{\bf{kq}}}^{\alpha \beta } = \left[ { - \frac{{{f_{2{\bf{k}}}}{f_{3{\bf{k}}}}}}{{{\eta _{\bf{k}}}{\mu _{\bf{k}}}}}\frac{{i{f_{1{\bf{q}}}}}}{{{\mu _{\bf{q}}}}} + \frac{{{f_{1{\bf{k}}}}{f_{3{\bf{k}}}}}}{{{\eta _{\bf{k}}}{\mu _{\bf{k}}}}}\frac{{i{f_{2{\bf{q}}}}}}{{{\mu _{\bf{q}}}}}} \right],\Phi _{{\bf{kq}}}^{\beta \alpha } = \left[ {\frac{{i{f_{1{\bf{k}}}}}}{{{\mu _{\bf{k}}}}}\frac{{{f_{2{\bf{q}}}}{f_{3{\bf{q}}}}}}{{{\eta _{\bf{q}}}{\mu _{\bf{q}}}}} - \frac{{i{f_{2{\bf{k}}}}}}{{{\mu _{\bf{k}}}}}\frac{{{f_{1{\bf{q}}}}{f_{3{\bf{q}}}}}}{{{\eta _{\bf{q}}}{\mu _{\bf{q}}}}}} \right],\\
\Phi _{{\bf{kq}}}^{\alpha \gamma } = \left[ {\frac{{{f_{2{\bf{k}}}}{f_{3{\bf{k}}}}}}{{{\eta _{\bf{k}}}{\mu _{\bf{k}}}}}\frac{{i{f_{2{\bf{q}}}}}}{{{\eta _{\bf{q}}}}} + \frac{{{f_{1{\bf{k}}}}{f_{3{\bf{k}}}}}}{{{\eta _{\bf{k}}}{\mu _{\bf{k}}}}}\frac{{i{f_{1{\bf{q}}}}}}{{{\eta _{\bf{q}}}}} - \frac{{{\mu _{\bf{k}}}}}{{{\eta _{\bf{k}}}}}\frac{{i{f_{3{\bf{q}}}}}}{{{\eta _{\bf{q}}}}}} \right],\,\,\Phi _{{\bf{kq}}}^{\gamma \alpha } = \left[ { - \frac{{i{f_{2{\bf{k}}}}}}{{{\eta _{\bf{k}}}}}\frac{{{f_{2{\bf{q}}}}{f_{3{\bf{q}}}}}}{{{\eta _{\bf{q}}}{\mu _{\bf{q}}}}} - \frac{{i{f_{1{\bf{k}}}}}}{{{\eta _{\bf{k}}}}}\frac{{{f_{1{\bf{q}}}}{f_{3{\bf{q}}}}}}{{{\eta _{\bf{q}}}{\mu _{\bf{q}}}}} + \frac{{i{f_{3{\bf{k}}}}}}{{{\eta _{\bf{k}}}}}\frac{{{\mu _{\bf{q}}}}}{{{\eta _{\bf{q}}}}}} \right],\\
\Phi _{{\bf{kq}}}^{\beta \gamma } = \left[ { - \frac{{{f_{1{\bf{k}}}}}}{{{\mu _{\bf{k}}}}}\frac{{{f_{2{\bf{q}}}}}}{{{\eta _{\bf{q}}}}} + \frac{{{f_{2{\bf{k}}}}}}{{{\mu _{\bf{k}}}}}\frac{{{f_{1{\bf{q}}}}}}{{{\eta _{\bf{q}}}}}} \right],\,\,\Phi _{{\bf{kq}}}^{\gamma \beta } = \left[ { - \frac{{{f_{2{\bf{k}}}}}}{{{\eta _{\bf{k}}}}}\frac{{{f_{1{\bf{q}}}}}}{{{\mu _{\bf{q}}}}} + \frac{{{f_{1{\bf{k}}}}}}{{{\eta _{\bf{k}}}}}\frac{{{f_{2{\bf{q}}}}}}{{{\mu _{\bf{q}}}}}} \right].
\label{eq:HUp_Psi}
\end{multline}
\endlongformula
The most significant values of the intra-cluster structural factors ${\Phi _{000}^{{\rho _i}{\rho _j}}}$:
\begin{equation}
\Phi _{000}^{\alpha \alpha }=0.6159,\,\,\,\Phi _{000}^{\beta \beta }=0.9179,\,\,\,\Phi _{000}^{\gamma \gamma }=0.9159
\end{equation}
Each term of the intra-atomic Coulomb interactions in terms of oxygen molecular orbitals has the operator structure ${\rho _{i{\bf{f}}\sigma }^\dag {\rho _{i'{\bf{f'}}\sigma }}\rho _{j{\bf{g}}\sigma '}^\dag {\rho _{j'{\bf{g'}}\sigma '}}}$ consisting of all possible products of four operators each of which creates or annihilate a hole in one of the three possible oxygen molecular orbitals in different clusters. Here are the values of some intra-cluster coefficients: $\Psi _{0000}^{\alpha  \uparrow \alpha  \uparrow \alpha  \downarrow \alpha  \downarrow }=0.7447$, $\Psi _{0000}^{\beta  \downarrow \beta  \downarrow \beta  \uparrow \beta  \uparrow }=0.21$, $\Psi _{0000}^{\gamma  \downarrow \gamma  \downarrow \gamma  \uparrow \gamma  \uparrow }=0.178$, $\Psi _{0000}^{\alpha  \downarrow \alpha  \downarrow \beta  \uparrow \beta  \uparrow }=0.018$, $\Psi _{0000}^{\alpha  \downarrow \alpha  \downarrow \gamma  \uparrow \gamma  \uparrow }=0.0324$.
%\longformula
%\endlongformula

\end{multicols}\twocolumn

%\end{multicols}{2}
%\end{multicols}\twocolumn

\end{document}